\documentclass{aa}

\usepackage{graphicx}
\usepackage{epstopdf}
\usepackage{threeparttable}

\usepackage{txfonts}

\usepackage{ulem}

\usepackage{xcolor}

\usepackage{soul}
\usepackage{graphicx}
\usepackage{subcaption}

\usepackage{float}

\usepackage{array, multirow, bigdelim, makecell, booktabs} 

\usepackage{threeparttable}

\usepackage{lscape}
\usepackage[switch]{lineno}

\usepackage{longtable}

\usepackage[colorlinks = true, linkcolor=blue, citecolor = blue]{hyperref}
\begin{document} 

\title{Cataclysmic Variable Candidates Identified in eROSITA-DE DR1, XMM-Newton, Swift, and ROSAT Catalogs}

\titlerunning{Candidate of  Cataclysmic Variable in four X-ray catalogs}


   \author{Xiangxiang Wang
          \inst{1}
          \and
          Jumpei Takata\inst{1}
          }

   \institute{$^1$ Department of Astronomy, School of Physics, Huazhong 
     University of Science and Technology, Wuhan, Hubei 430074, China \\   
   \email{wangxx@hust.edu.cn,takata@hust.edu.cn} 
             }

   \date{Received xx; accepted xx}


  \abstract
      {
        The cataclysmic variable (CV) is a binary system composed of a white dwarf (WD) primary and a low-mass main-sequence star. The accretion onto the WD surface causes an X-ray emission, and the CVs are one of the major Galactic X-ray sources. The eROSITA observation offers new opportunity to identify numerous X-ray emitting CVs and advance the understanding for its Galactic population.}
  {We carry out searching for candidates of the X-ray emitting CVs in the four catalogs of eROSITA, XMM-Newton, Swift and ROSAT observations. We also examine how the eROSITA survey can enhance the search for new X-ray emitting CVs.
   }
  {In this study,  we identify the candidates of the X-ray emitting CVs by cross-matching  between the four X-ray catalogs and the GAIA sources that are selected from the  bridge region between the main-sequence and the WD’s cooling-sequence in the Hertzsprung-Russell diagram. For the selected candidates, we search for the periodic modulation in the photometry light curves taken by the ZTF and/or TESS observations.
   }
  {
The cross-matching selects 444 candidates of the X-ray emitting CVs, in which 267 sources are the identified CVs.  In other 177 candidates of the X-ray emitting CVs, we identify possible orbital modulation for 56 sources in  ZTF/TESS light curves.
    About 51\% of the candidates are exclusively selected from the eROSITA catalog, while about 36\%, 9\% and 4\% are selected from the XMM-Newton, Swift and ROSAT catalogs, respectively. On the color-color diagram, most of the selected CV candidates are discriminated from the active stars, while some sources are likely classified as eclipsing binary systems 
    rather than CVs.}
{
Combining the cross-matching process between X-ray catalogs and GAIA sources with the timing analysis of the ZTF/TESS data is an effective tool to identify the candidates of X-ray emitting CVs. In particular, the eROSITA observation has successfully identified the sources that have been missed in other three catalogs, because of its wide coverage on the sky region and high flux sensitivity. On the other hand, the current flux sensitivity of the eROSITA is limited to the order of  $\sim 10^{-14}~{\rm erg~cm^{-2}~s}$. As  the first  eROSITA catalog is result of the operation in the first six months, it will be expected that the future operation will further increase the population of the X-ray emitting CVs.}


   \maketitle
%
\section{Introduction} \label{sec:intro}
White dwarf (hereafter WD) is common endpoint of the stellar evolution
for $>90$\% of the main-sequence star. The cataclysmic variable (hereafter CV) is a binary system composed of a WD primary and a low-mass main-sequence star \citep{1995cvs..book.....W}, and the orbital period is usually less than one day. The secondary fills its Roche-lobe, and hence the CVs have been known as a binary system  operating accretion process. The CVs are usually observed with an transient optical outburst (dwarf nova) due to the disk instability and show a steady X-ray emission  due to the accretion onto the WD surface. Because of its large population, the CVs are one of the major X-ray sources in the Galaxy.  

Increasing in the population of the X-ray sources provides an opportunity
to carry out identification for new CVs by cross-matching between the X-ray source catalogs and optical source catalogs. An X-ray survey for the CVs is important for  identifying the magnetic CVs \citep[e.g.,][]{2020NewAR..9101547L}, for which the optical surveys alone  may be difficult and inefficient to be identified because of  their lack of frequent outburst behavior. The increasing in the CVs identified in X-rays will overcome the bias of the optical survey in the population of the CVs.  Moreover, identifying the X-ray emitting CVs will also help us to exclude the CV-candidates from the list of the candidate of new type of the compact X-ray sources, such as the emission from the isolated black hole in the Galaxy \citep{2021NatCo..12.5615K}. 

Dedicated studies for identifying the X-ray emitting CVs have been conducted  in previous studies. For example, \cite{2014MNRAS.441.1186D} search for the CVs from the Catalina Real-time Transient Survey and identify 855 CV-candidates. They cross-match their CVs with ROSAT source catalog and find 42 X-ray emitting CVs or its candidates.
\cite{2022ApJ...936..134T} carry out the cross-matching between X-ray sources in the catalogs  (ROSAT, Swift and XMM-Newton)  and the GAIA sources that are located in the bridge region  between the main-sequence and the WD's cooling-sequence in the Hertzsprung-Russell diagram. They determine the orbital characteristics for eight X-ray CV-candidates using the photometric light curves in the optical bands.
\cite{2024A&A...690A.374G} search for new CVs in Chandra Source Catalog v2.0, cross-matching  with Gaia data-release 3 (DR3). They report 14 new CV-candidates and confirm the CV nature of four CV-candidates in the follow-up optical observations.
These results show that the conduct of a joint  X-ray and optical studies is a useful tool for finding new CVs and its candidates.

Extended Roentgen Survey with an Imaging Telescope Array
  (hereafter eROSITA) and Mikhail Pavlinsky Astronomical Roentgen Telescope–X-ray Concentrator (ART-XC telescope)
  on the Spectrum-Roentgen-Gamma (SRG) mission began new all sky survey in 2019 \citep{2021A&A...656A.132S}.
\cite{2022A&A...661A..39Z} identify three X-ray emitting CVs using eROSITA and ART-XC telescope with optical follow-up observations.
\cite{2022A&A...661A..43S} report the discovery of an eclipsing polar, eRASSt J192932.9-560346/Gaia21bxo, identified through  SRG/eROSITA and Gaia transient surveys. They also confirm the orbital period of this polar to be $\sim 92.51$~minutes. \cite{2023A&A...676A...7M} use the XMM-Newton and eROSITA observations to confirm the X-ray emissions from three candidates of the  period-bounce CVs (V379~Vir, SDSS J151415.65+074446.5 and SDSS J125044.42+154957.4) with a luminosity of $L_X\sim 10^{29}~{\rm erg~s^{-1}}$, suggesting the presence of the accretion process in these binary systems. \cite{2023ApJ...954...63R} conduct follow-up optical observation for SRGeJ045359.9+622444, which is a new eclipsing AM CVn system, discovered from  joint eROSITA and  Zwicky Transient Facility \citep[hereafter ZTF,][]{2019PASP..131a8003M} observations to identify the CVs, and confirm the orbital period of $\sim 55.08$~minutes. Finally, \cite{2024MNRAS.528..676G} also utilize the joint eROSITA and ZTF search and discover an eclipsing CV (SRGeJ041130.3+685350). These studies demonstrate the potential of eROSITA observations to identify new X-ray emitting CVs.

The X-ray source catalogs of the observation by  SRG observatory  enable us an extensive search for new CVs  and its candidates.
\cite{2023ApJ...945..141R} carry out cross-matching between Final Equatorial Depth Survey (eFEDS) sources of eROSITA  and the photometry catalog of the ZTF observation,  and they identify two new polars. \cite{2024A&A...686A.110S} also use the catalog of eFEDS survey and identify 26 CVs through the follow up observation by Sloan Digital Sky Survey.  \cite{2024A&A...687A.183S} report the catalog of 469 Galactic sources measured by SRG/ART-XC telescope, in which 192 sources belong to previously identified CVs or the candidates of CVs. \cite{2025PASP..137a4201R} carry out cross-matching between eROSITA-DE Data Release 1 (DR1) and GAIA~DR3 and identify X-ray emitting CVs located within 150~pc. They estimate that the number density of the CVs  in the solar neighborhood is  $\rho_N\sim (3.7\pm 0.7)\times 10^{-6}~{\rm pc}^{-3}$. Finally,  \cite{2024A&A...690A.243S} reports a characteristic of $\sim 400$ CVs and its candidates selected from eROSITA-DE DR1, which significantly increases the population of CVs identified by the X-ray observations.

The sky survey of eROSITA is about 25 times more sensitive than the previous ROSAT all-sky survey \citep{2021A&A...647A...1P} and covers a Galactic hemisphere.
With its high sensitivity and large covering sky region, the eROSITA survey offers a new opportunity to identify numerous X-ray emitting CVs. In this study, therefore,  we aim to search for the X-ray emitting CVs and its candidates in the eROSITA source catalog.  We select the eROSITA sources by cross-matching with the GAIA sources that are potential candidates of CV based on the color–magnitude diagram. This cross-matching process is similar to study done by \cite{2025PASP..137a4201R}. In our study, we further search for a periodic signal in photometric data of the selected targets taken  by ZTF and The Transiting Exoplanet Survey Satellite \citep[hereafter TESS,][]{2015JATIS...1a4003R}. We also study how the eROSITA survey enhances the identification for the X-ray emitting  CVs. In Sect.~2, we describe the data reduction and method to search for the periodic modulation in the optical light curve. In Sect.~3, we present the results of the survey for the X-ray emitting CVs. We summarize our results in Sect.~4.

 \section{Sample selection and analysis process}
In this section, we present the sample selection and process to identify the candidates of the X-ray emitting CVs. An outline of the process is as follows:
\begin{enumerate}
    \item Select the GAIA sources from the bridge region between main-sequence and the WD's cooling sequence in Hertzsprung-Russell diagram (Sect. 2.1). 
\item Cross-matching between the selected GAIA sources and eROSITA catalog (Sect. 2.2).
\item Timing analysis of the light curves measured by ZTF and TESS observations (Sects. 2.3 and 2.4).
\item Evaluation of the possibility of other type X-ray sources (Sects. \ref{sec:xpro} and \ref{sec:dif}).
\end{enumerate}

\subsection{GAIA data}
\label{sec:gdata}
The GAIA DR3 contains  about 1.8 billion sources observed by the $G$-band photometry and 
about 1.5 billion sources with $G_{\rm BP}$ and $G_{\rm RP}$ band photometry, complemented by positions on the sky, parallax, proper motion, etc \citep{2023A&A...674A..14R}. We download the data from the GAIA Archive\footnote{\url{https://gea.esac.esa.int/archive/}} using \verb|astroquery| \citep{2019AJ....157...98G} .

\begin{figure}
\includegraphics[scale=0.6]{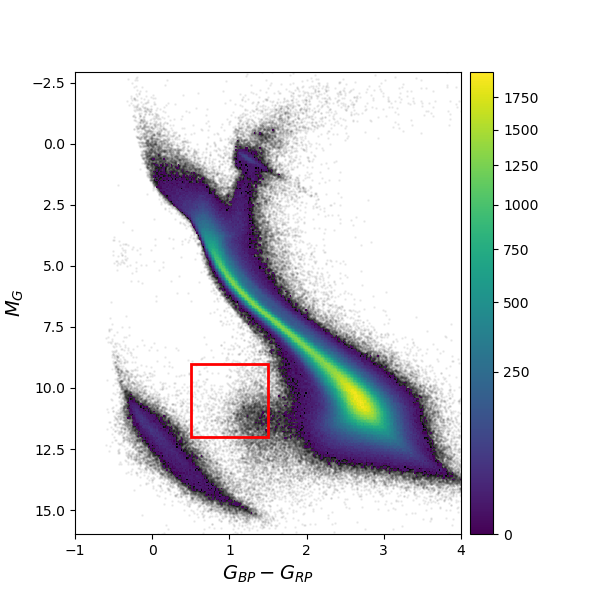}
\caption{Hertzsprung-Russell diagram with the GAIA sources. The red square indicates the region 
from which we select the GAIA sources.} 
\label{fig:general}
\end{figure}

It has been known that many CVs locate in the region between the main-sequence and WD's cooling sequence on the Hertzsprung-Russell diagram. We, therefore, select the GAIA's source by limiting (i) a color as $0.5<G_{\rm BP}-G_{\rm RP}<1.5$, where $G_{\rm BP}$ and $G_{\rm RP}$ are blue and red magnitude defined by the GAIA photometric system, respectively, and (ii) a magnitude of $9<M_{\rm G}<12$, as indicated in 
Fig.~\ref{fig:general}: we limit the search region to obtain manageable sample size.  
We apply additional conditions in order to select clean samples \citep{2018A&A...616A...2L}:
\begin{enumerate}
    \item \verb|parallax_over_error > 5|     
\item \verb|phot_bp_mean_flux_over_erro > 8|
\item \verb|phot_rp_mean_flux_over_error > 10|
\item \verb|astrometric_excess_noise < 1|
\item \verb|phot_bp_rp_excess_factor < 2.0+0.06*|
\verb|power(phot_bp_mean_mag-phot_rp_mean_mag,2)|
\item \verb|phot_bp_rp_excess_factor > 1.0+0.015*|
 \verb|power(phot_bp_mean_mag-phot_rp_mean_mag,2)|
\item \verb|visibility_periods_used > 5|.
\end{enumerate}
We obtain a list of 121,488 GAIA sources.

\subsection{Cross-matching with eROSITA catalog}
\begin{figure}
\includegraphics[scale=0.5]{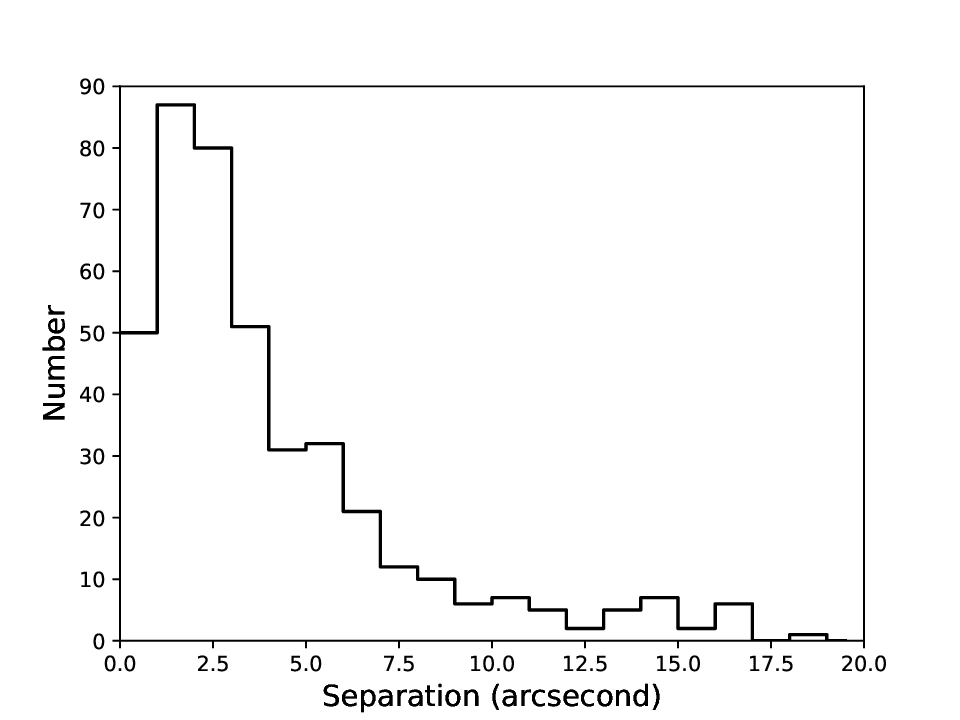}
\caption{Distribution of the angular separation between the eROSITA position and the GAIA position for the identified CVs. The data is taken from \cite{2024A&A...690A.243S}. } 
\label{fig:sepa}
\end{figure}
The eROSITA-DE Data Release 1 (DR1) comprises data from the first six months of the SRG/eROSITA all-sky survey (hereafter eRASS1) 
and it includes 930,203 sources located at the Galactic hemisphere \citep{2024A&A...682A..34M}. We carry out the cross-matching between the sources in the eROSITA catalog and the 121,488  GAIA sources that are described in Sect.~\ref{sec:gdata}. We select the closest GAIA source located within a 10” radius, which is the typical angular resolution of the eROSITA observation \citep{2021A&A...647A...1P,2024A&A...682A..34M},  from the center of the GAIA sources. We also check the distribution of the angular separation between eROSITA position and GAIA position for the identified CVs presented in \cite{2024A&A...690A.243S}. As indicated in Fig.~\ref{fig:sepa}, more than 90~\% of the sources have an angular separation smaller than 10''. Our cross-matching process selects 264 GAIA/eROSITA sources.

We cross-match the selected GAIA/eROSITA sources with the SIMBAD astronomical database\footnote{\url{https://simbad.cds.unistra.fr/simbad/}} \citep{2000A&AS..143....9W}, the International Variable Star Index \footnote{\url{https://www.aavso.org/vsx/index.php}}  (hereafter VSX), the Open Cataclysmic Variable Catalog~\footnote{\url{https://depts.washington.edu/catvar/index.html}} (hereafter OCVC) and  Ritter Cataclysmic Binaries Catalog \footnote{\url{https://heasarc.gsfc.nasa.gov/W3Browse/star-catalog/rittercv.html}} \cite[hereafter RK,][]{2003A&A...404..301R}, which list numerous CVs and CV-candidates. We divide the selected GAIA/eROSITA sources into two categories: (i) identified CVs, which are sources listed as CVs in SIMBAD,  VSX,  
 OCVC and RK databases, and (ii) others, which are referred as CV-candidates in this study (Table~\ref{table:cross}). We  note that many VSX-listed sources are discovered by an optical transient survey, for example, Catalina Real-Time Transient Survey \citep{2009ApJ...696..870D} and All-Sky Automated Survey for SuperNovae~\citep{2014ApJ...788...48S}. Hence, our CV-candidates listed in VSX are more likely CVs. Additionally, some of  sources in  CV-candidates are listed as WD in the primary or secondary category of the SIMBAD database, suggesting the accreting WDs candidates. 

We note that in our cross-mathcing process with 121,488 GAIA sources, two eROSITA sources have multiple (two) GAIA sources located within 10'' search radius. 1eRASS J170454.8-291337, which is categorized as a CV-candidate, has potential Gaia counterparts DR3 6029819936903977856 and  6029819936840722048, which are separated by 7.53'' and 7.69'',  respectively, from the eROSITA source. The latter GAIA source is a Young Stellar Object Candidate in SIMBAD. Consequently, we cannot exclude the possibility that the eROSITA is the X-ray counterpart of this GAIA source. 1eRASS J180018.7-353311 (CV) also has two counterparts, Gaia DR3 4041983178121822976 and 4041983178121844224, which are separated by  4.60'' and 5.22'' respectively, both of them correspond to TCP J18001854-3533149, a identified CV in VSX. Since Gaia DR3 4041983178121844224 is more close to TCP J18001854-3533149, we anticipate this Gaia source as the counterpart of this eROSITA source.  In addition, if we cross-match our eROSITA sources with all GAIA DR3 sources, 153 eROSITA sources have multiple GAIA sources with on average  6 possible counter parts; the maximum is 37 GAIA sources.  Hence despite the expected X-ray luminosity and the optical to X-ray flux ratios of most CV-candidates selected in this study are consistent with identified CVs (sections~\ref{sec:xpro}), multi-wavelength follow-up observations are necessary to confirm their nature. 

\subsection{ZTF observation}
\label{sec:ztf}
\begin{figure}
\includegraphics[scale=0.65]{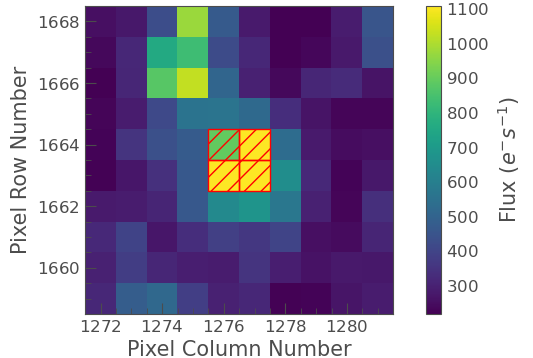}
\caption{TESS full-frame images from sector 39 for the region around  GAIA DR3 5822540653269409408. The LS-periodogram is created for each pixel in the figure, and  the periodic signal detected only pixels around the target (indicated by red color square) is considered as a possible signal from the target. }
\label{fig:pixel}
\end{figure}

\begin{figure*}
\includegraphics[scale=0.55]{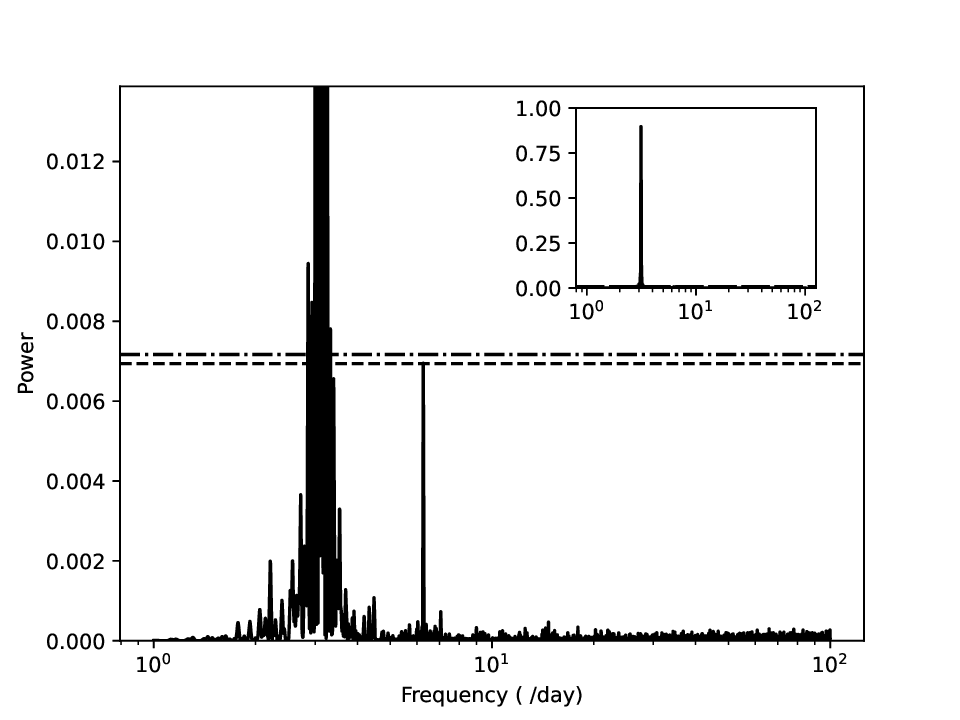}
\includegraphics[scale=0.55]{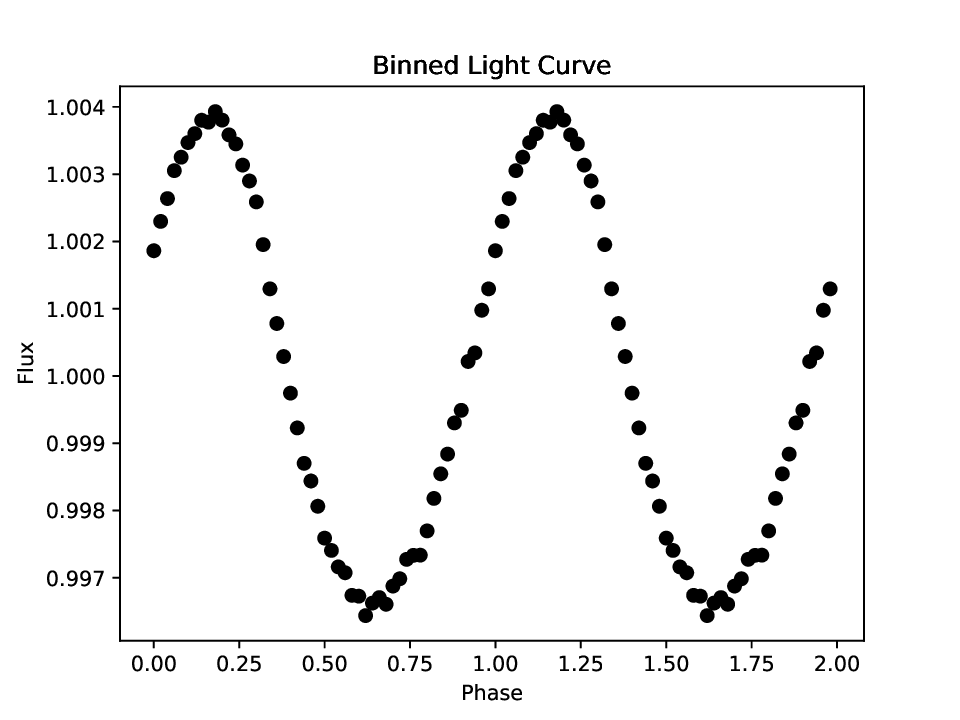}
\caption{LS periodogram (left panel) and the folded light curve (right panel) with the TESS data for GAIA DR3 5822540653269409408, which is selected as the counterpart of 1eRASS J155030.1-654403. The black dashed line and the black dashed-dotted line are FAP=0.01 determined by the methods of 
\cite{2008MNRAS.385.1279B} and of the bootstrap  \citep{2018ApJS..236...16V}, respectively. The LS diagram reveals $\sim3.13(3)~{\rm day^{-1}}$ modulation and its first harmonic. Two period cycles of the light curve is presented for clarity. }
\label{fig:LS}
\end{figure*}
One of the characteristic of the CVs is the orbital modulation in the optical photometric light curve caused by an eclipse of one star by another or the elliptical shape of the companion star or the heating of the surface of the companion star by irradiation from the WD or disk.  Another characteristic is  outbursts (dwarf nova) due to an instability of the accretion disk of non-magnetic CVs. We focus on the searching for CV-candidates by identifying the possible signal of the orbital modulation in the optical light curves and confirm the binary nature of our CV-candidates. To perform  a timing analysis, the cadence of the  GAIA  observations are insufficient to search for the periodic modulation in the light curves. In this study, therefore, we use the photometric light curves taken by ZTF\footnote{\url{https://irsa.ipac.caltech.edu/frontpage/}}~\citep{2019PASP..131a8003M} and/or  TESS\footnote{\url{https://mast.stsci.edu/portal/Mashup/Clients/Mast/Portal.html}} \citep{2015JATIS...1a4003R}.

We cross-match the selected eROSITA/GAIA sources with the  ZTF DR-21 objects and  search for a periodic modulation in the photometric light curve. We create a Lomb-Scargle periodogram~\citep[hereafter LS,][]{1976Ap&SS..39..447L} using  the python package \verb|LombScargle| in \verb|astropy| \citep{2022ApJ...935..167A}.  We evaluate the  false alarm probability of the periodic signal with the methods of \cite{2008MNRAS.385.1279B} and of the bootstrap \citep{2018ApJS..236...16V}. We find that the ZTF observations for most of our targets provide only several hundred data points, and the quality of the light curve is insufficient to evaluate the periodic signal. Additionally, a significant portion of our targets are located outside the field of view of the ZTF observations. 
The ZTF observation covers only 50 sources among $\sim 180$ CV-candidates, in which 34 sources have enough number (>200) of the data points for timing analysis. Hence, conducting a detailed search for periodic signals using ZTF data is challenging for most cases. For periodic presented in this study, we denote the uncertainty of the frequency estimated  from  the Fourier resolution of the observation (namely, the inverse of the time span covered by the observation).

\subsection{TESS observation}
Because of the space-based observation, TESS covers a large area of the sky and provides the photometric data for the region around most of the targets examined in this study. One sector of the TESS observations provides the light curve of about 27-days long, which enables us a detailed investigation for the periodic modulation of a timescale of one day. Moreover, the different sectors of the TESS observations have covered one sky region, enabling us to discriminate true periodic signal from spurious one.  On the other hand, the angular resolution  covered by one pixel is of the order of 20''. With such a insufficient angular resolution, one pixel may contain several optical sources. Consequently,  even when a periodic signal is detected in the photometric light curve corresponding to a pixel that includes our target, we cannot definitively attribute the signal to the target itself.

For each target, we examine
all available TESS light curve files and/or full-frame images (hereafter FFIs). We utilize the TESS analysis tool \verb|Lightkurve| \citep{2018ascl.soft12013L} to analyze the FFIs, and cutout the image into $10\times 10$ pixels (Fig.~\ref{fig:pixel}).  We extract the light curve information of each pixel to search for the periodic modulation in LS-periodogram: for FFIs, we are unable to  extract the meaningful light curves from some pixels due to the insufficient quality of the data (e.g., NAN flux value). To enhance the likelihood that the observed signal is associated with our target, we apply the following criteria:

\begin{enumerate}
\item  The LS power should be at least double the power associated with a False Alarm Probability (FAP) of 0.01. This will ensure that the signal stands out significantly against the background noise.

\item The signal is detected from only  pixel around the target (shaded region in Fig.~\ref{fig:pixel}). This may avoid the possibility that the detected signal originates from either contamination of nearby brighter source or the spurious signal.
\end{enumerate}

Figure~\ref{fig:pixel} and Figure~\ref{fig:LS} present an example of identifying a CV candidate using TESS data.  
Figure~\ref{fig:pixel} shows the TESS FFI  from sector 39 centered on the source GAIA DR3 5822540653269409408, which is selected as an optical counterpart of  1eRASS J155030.1-654403. For each pixel in the figure, we create LS-periodogram in a frequency range of 
 1–100~day$^{-1}$, which corresponds to typically period of the observed CVs. In the LS-periodogram for the light curve extracted from  the central region in FFIs, we identify a strong period signal at $\sim 3.13(3)~{\rm day}^{-1}$, as shown in Fig.~\ref{fig:LS}, with LS-power higher than twice of the corresponding to FAP=0.01.  We also identify  the periodic signal using FFIs from  sectors 12 , 65 and 66, suggesting this periodic signal  is likely related to an astrophysical source.  If the periodic signal is indeed related to the GAIA source, the orbital signal of the accreting WDs candidates will be the most natural explanation, since the GAIA source is selected from the bridge region between the main-sequence and WD's cooling sequence in the H-R diagram (Fig.~\ref{fig:general}). We classify this source into the CV-candidate, since no record of this source is found in SIMBAD, VSX , OCVC and RK databases. The parallax measured by GAIA suggests the distance to this source is 570 pc. The eROSITA observation suggests that the luminosity in $0.2-8.0$~keV of this source is $\sim 1.6 \times 10^{30}~{\rm erg~s^{-1}}$, which is also a typical value for CVs. Finally, if we cross-match the eROSITA source with the whole GAIA sources, the 1eRASS J155030.1-654403 has five GAIA sources within the radius of 10''. None of those GAIA sources are listed in 
 SIMBAD and other databases used in this study.
 
We note that each target has been normally covered by the several sectors with different readout times of the TESS observations. This helps us to discriminate true signal from an aliasing signals caused by the effect of the readout time.  If the periodic signal is confirmed only at one sector, as described below, we cannot discriminate between true signal and the aliasing signal. In such cases, we report a longer one, which is usually on the order of hours and is more close to typical orbital period of the CVs.  We find that several sources show 
the significant periodic signal only at one sector but no signal in other sectors. One possibility related to such a transient periodic signal is a superhump, which is a periodic variation in the emission observed from an eccentric disk after the outburst of CVs \citep{1995cvs..book.....W}.  For example, 3 sectors of the TESS observations cover the region of GAIA 5710755475028251776/1eRASS J075656.3-231557, but only data of sector~34 shows significant modulation with a period of $\sim 0.0892(4)$~days. This period is likely period of the superhump after the  outburst happen in early 2021.  CBA Extremadura
Observatory observation carried out at 2021 February 28th also confirm the period of the superhump of  0.0970 $\pm$ 0.0027~day \footnote{\url{http://ooruri.kusastro.kyoto-u.ac.jp/mailarchive/vsnet-alert/25465}}, which is close to value of the TESS observation.  Since the superhump period typically varies by a few percent compared to the orbital period, we regard the observed modulation could indicate the actual orbital period.

\begin{figure*}[t]
\includegraphics[scale=0.5]{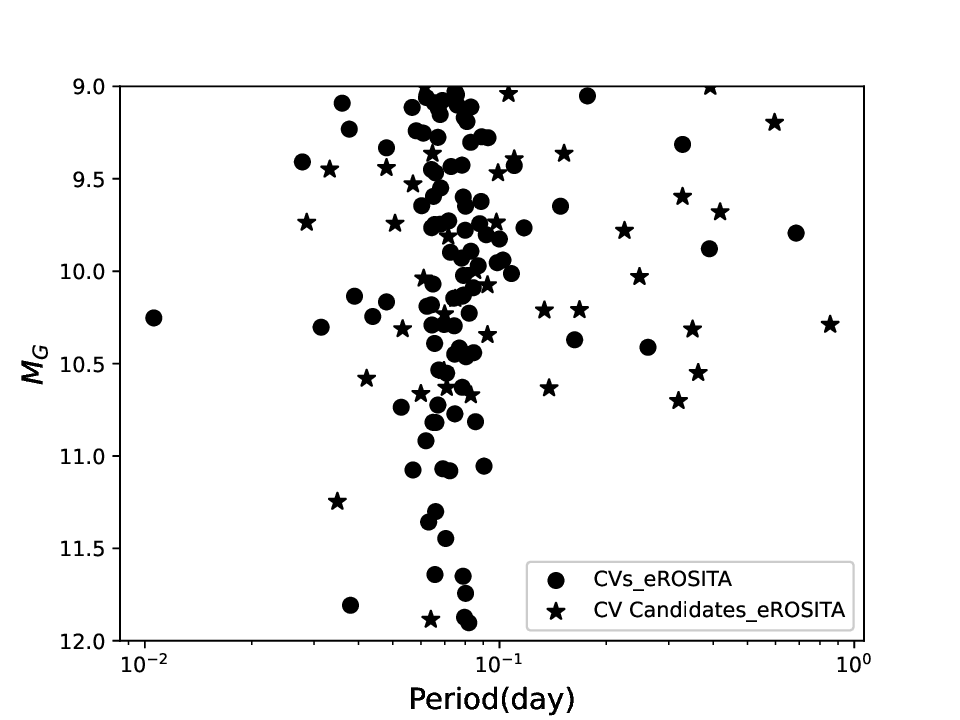}
\includegraphics[scale=0.5]{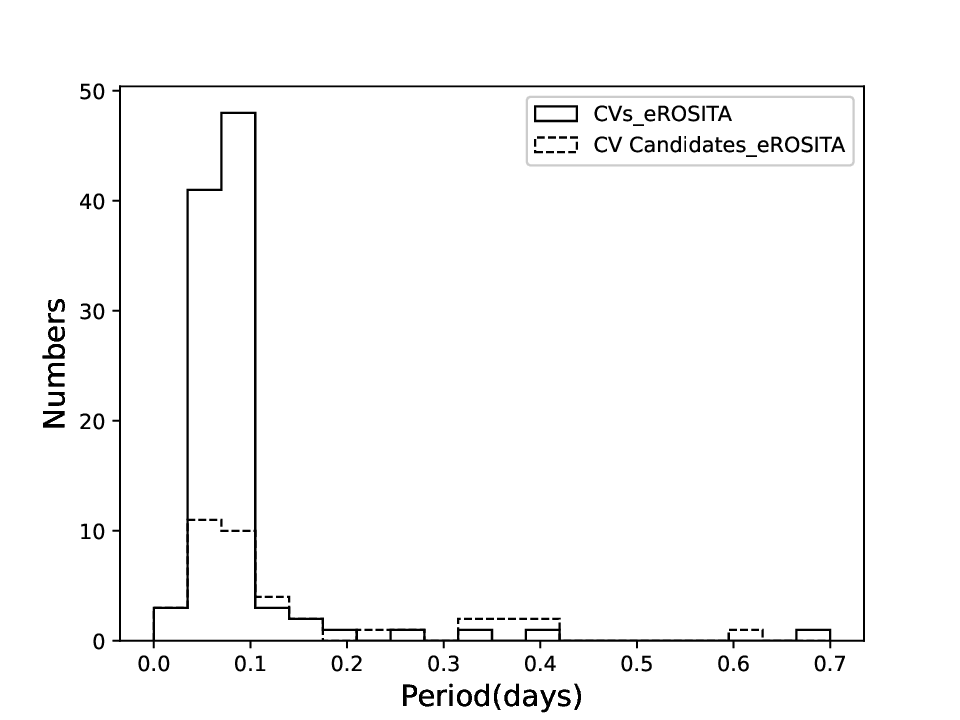}
\caption{Periodic signal from CVs (102 sources, filled circles) and CV-candidates (40 sources, stars) selected by the  cross-matching with eROSITA catalog. Left: GAIA G-band magnitude vs. observed period of the modulation in ZTF and/or TESS light curves.  Right: Distribution of the observed periods.}
\label{fig:period-mg}
\end{figure*}

\subsection{Other X-ray observations}
\label{otherx}
We also use the X-ray catalogs of the XMM-Newton \citep[DR13,][]{2020A&A...641A.136W}, Swift \citep[Second Swift-XRT Point Source (2SXPS) Catalog,][]{2020ApJS..247...54E}, and ROSAT \citep[Second ROSAT all-sky survey (2RXS) source catalog,][]{2016A&A...588A.103B}. The  error on the position of the XMM-Newton observations can be smaller than a few arcseconds for on-axis observations, but exceeds 10'' for off-axis observations. The typical position errors of the 2SXPS catalog and 2RXS catalog are 5.6'' and greater than $\sim7"$, respectively. To maintain consistency with the cross-matching process of the eROSITA catalog, 
we select the X-ray sources located within 10'' at the center of the GAIA's position. Through a cross-matching with the  GAIA sources, we select 109 XMM-Newton sources, 111 Swift sources and 69 ROSAT sources.  Out of these, 44 from XMM-Newton, 91 from Swift, and 59 from ROSAT are identified CVs (Table~\ref{table:cross}). As we expect, the eROSITA observation alone can identify more identified CVs and its candidates  compared to other X-ray observations.

\begin{table}
\centering
\caption{Cross-matching for X-ray catalogs with 121,488 GAIA sources}
\begin{tabular}{ccccc}
\hline
Telescope &  Total & CV$^{\rm a}$  &  CV-candidates$^{\rm b}$ &Duplicates$^{\rm c}$ \\
\hline
eROSITA & 264 & 173 & 91~(40$^{\rm d}$)&5 \\
XMM-Newton  & 109  & 44 & 65~(11)&4\\
Swift & 111  & 91& 20~(6)& 5\\
ROSAT & 69  & 59 & 10~(3)& 3 \\
\hline
\end{tabular}
\begin{tablenotes}
\item{a}: Identified CVs in SIMBAD, VSX , OCVC and/or RK.
\item{b}: Sources excluding the identified CVs.
\item{c}: Number of sources that are detected in multiple catalogs for CV-candidates.
\item{d}: Number of sources with a detection of periodic signal in ZTF and/or TESS light curves.
\end{tablenotes}
\label{table:cross}
\end{table}

\begin{table*}
\caption{CVs found in all four catalogs.}
\label{table:unique CVs}
\centering           
\begin{tabular}{llll}
\toprule
GAIA ID &Name & Frequency  & Type \\
& & (day$^{\rm -1}$)$^{\rm a}$& \\
\hline
6453536224527716224 & CD Ind&  13.12(2)  & Polar  \\
674214551557961984 & U Gem&  5.65(2)  & Dwarf novae  \\
3116059834803771904 & PM J07068+0324&  14.10(2)  & Polar  \\
6050296829033196032 & AR Sco&  6.7324(4)  & Eclipsing CV  \\
\hline
\end{tabular}
\begin{tablenotes}
\item{a}: Orbital frequency found in ZTF/TESS data.
\end{tablenotes}
\end{table*}

\begin{table*}
  \caption{ZTF sources with the periodic modulation in CV-candidates.}
\label{table:ztf}
  \centering
  \resizebox{\textwidth}{!}{
  \begin{tabular}{lllllll}
    \toprule
GAIA ID & Name$^{a}$ &X-ray source & R.A. & DEC.  & Frequency &  Remarks$^{b}$\\
&  & & (degree) & (degree) &(days$^{-1}$)& \\
\hline
 3208641703155603584 && 1eRASS J051248.0-055001 & 78.200 & -5.835&1.6756(4)& TESS \\ 
3348188519201136896 & & 1eRASS J054726.8+154051 &86.861& 15.680&  31.0357(4)$^{c}$& TESS \\
6286158119187026688 &HE 1432-1625&1eRASS J143545.7-163819 &218.939&-16.838 &2.8559(4)&  VSX\\
6278962983814509184 &HE 1436-2137& 4XMM J143912.6-215014
 & 219.802& -21.837& 0.4585(4) & TESS \\
406056486024031872 && 2SXPS J014816.9+510117 &27.071 & 51.021&   11.2291(4)& TESS
\\
4299973372442437760 &ZTF J200023.54+091547.6 &2SXPS J200023.5+091547 & 300.098 & 9.263& 16.2336(4) &VSX\\
\hline
\end{tabular}
}
\begin{tablenotes}
\item{a}: Name listed on SIMBAD and/or VSX database.
\item{b}: Confirmation of the periodic signal in TESS data or VSX database.
\item{c}: ZTF data shows the periodic signal at the first harmonics. 
\end{tablenotes}
\end{table*}

\begin{table}
\centering 
\caption{Results of timing analysis with TESS light curves  for identified CVs.}
\label{table:tess}
\begin{tabular}{llllll}
\hline
Telescope &  Total & $N_{P_0}$  & $N_{P_1}$ & $N_{oth}$ & $N_{non}$  \\
\hline
Schwope et al (2024) &401 & 140 & 9 & 6  & 38\\
\hline
eROSITA & 173 & 30 & 12 & 2 & 35\\
XMM  & 44 & 20 & 6 & 1 &7 \\
Swift &91 & 25 &16 & 3 & 12 \\
ROSAT &59 & 18 & 6 & 1  & 10\\
\hline
\end{tabular}
\begin{tablenotes}
\item{$N_{P_0}$}: Signal at the orbital period. 
\item{$N_{P_1}$}: Signal at the first harmonics. 
\item{$N_{oth}$}: Signal that may be unrelated to the orbit period reported in the VSX database. 
\item{$N_{non}$}: Signal detection in ZTF/TESS but not reported in VSX. 
\end{tablenotes}
\label{table:CV}
\end{table}



\section{Results}
We select 121,488 GAIA that are selected with the conditions described in Sect.~\ref{sec:gdata}.  Table~1 shows the results of the cross-matching for four X-ray catalogs with the selected GAIA sources. For example, we select 264 eROSITA sources as the potential X-ray counterparts, which include  173 identified CVs and  91 CV-candidates. Avoiding duplication, we select 444 X-ray sources from the four catalogs as  potential counterparts to  our selected GAIA sources.
For the CV-candidates, we analyze  ZTF and TESS data to search for the periodic signal and present number of the sources with detection of the period signal in Table~1. 
For the eROSITA sources, for example, we identify the periodic signal from 40 sources categorized as the CV-candidates.
Among the 40 sources of CV-candidates, the periodic signal of four GAIA DR3 sources (5994940674888088832,~6286158119187026688, 5293314439454159488 and~ 4835425609500365184) have been reported in the VSX database.

\subsection{CVs covered by four catalogs} 
Among the selected GAIA sources in this study, 
there are four X-ray emitting CVs found across four catalogs, as presented in Table~\ref{table:unique CVs}. CD Ind (EUVE J2115–58.6) is a magnetic CV identified by \cite{1996IAUC.6297....2C}. \cite{1997A&A...326..195S} confirms it as a polar, but the WD's spin is not synchronized with the binary orbit. U Gem is a dwarf novae with eclipse feature \citep{2020MNRAS.492L..40A,1976ApJ...206..790A}. PM J07068+0324 (PBC J0706.7+0327) is detected with Swift/BAT and has been classified as  a polar due to its optical features, such as a variation of emission-line synchronized with the orbital phase \citep{2014A&A...561A..67P, 2015AJ....150..170H}. A spectroscopic observation conducted by the 2.4 m Hiltner telescope from the MDM Observatory finds an orbital period of 0.070907(11) day. We reconfirm this period, 0.0709(4) day, with the TESS observation. AR~Sco is an enclipsing CV, emitting across the electromagnetic spectrum from radio to X-rays wavelength bands,  with an orbital period of 3.56 hours. The brightness of the emissions varies with a period of 1.97 minutes, which is interpreted as the WD's spin period~\citep{2016Natur.537..374M}.

\subsection{Search for periodic modulation with ZTF data}
Due to  the insufficient  quality of the light curve and the limited field of view of the ZTF observation, a detailed searching for the periodic signal using 
ZTF data is  challenging for most of our targets. Among the CV-candidates in Table~\ref{table:cross}, we identify 6
sources with the periodic modulation in the ZTF data, as summarized in Table~\ref{table:ztf}.  The  identified
periods are consistent with the value found in TESS data and/or  reported in VSX database, as indicated in the Col. 7 of Table~\ref{table:ztf}.

Among the 6 sources in Table~\ref{table:ztf}, the eROSITA observation covers 3  sources. We identify two sources (GAIA DR3  3208641703155603584 and 406056486024031872) that are not listed in SIMBAD/VSX/OCVC/RK databases, and 4 sources for which their periodic modulations have not been reported in the VSX.  In Table~\ref{table:ztf}, the eROSITA catalog do not contain 3 ZTF sources. For 2 sources (2SXPS J014816.9+510117 and 2SXPS J200023.5+091547), the current eROSITA catalog does not cover their source positions.  For HE 1436-2137, its  X-ray flux, $\sim 2\times 10^{-14}~{\rm erg~cm^{-2}s^{-1}}$,  
may be too low to be detected with a half-year eROSITA observation (see Sect.~\ref{sec:comp}).

\subsection{Search for periodic modulation with TESS data}
Most of the periodic signals reported in this study are found by using the TESS data, which cover  about 156 sources in our CV-candidates. One issue to identify the source as the CV-candidate is that due to the limited spatial resolution of the TESS observation, we cannot claim whether the signal is related to the orbital motion, especially if there is no report of the periodicity in the VSX database.  Hence we evaluate the reliability that the detected signal is related to the orbital period of  the binary by analyzing the light curves of the identified CVs using TESS data. We analyze the TESS data for the CVs (i) selected from the eROSITA source catalog reported in  \cite{2024A&A...690A.243S} and (ii) selected from the four X-ray catalogs in our study (the sources in Col. 3 of Table~\ref{table:cross}).

\cite{2024A&A...690A.243S} present characteristics of the CVs selected from eROSITA-DE DR1. We check the TESS data of 401 CVs and CV-candidates, after removing sources categorized as the symbiotic stars,  listed in their list. 
The results of the timing analysis of the TESS data are summarized in the second line of Table~\ref{table:tess}. Among 401 sources, 140 sources exhibit TESS periods that align with those listed in VSX database ($N_{P_0}$ in Table~\ref{table:tess}), while 9 sources display the strongest signal  corresponding to the first harmonics ($N_{P_1}$). We have 6 sources ($N_{oth}$) from which the periodic signal confirmed with the TESS data is different from that reported in the VSX database. We also obtain $N_{non}=38$ sources that have a periodic signal in the TESS light curves but without recorded in the VSX database.
For those 38 sources with the period detection, we also check the RK catalog, and find the information of the periods for the six sources, among which  the periods of the five sources  are consistent with  the periods detected in  the TESS data. Our results would suggest that the TESS data are useful in identifying the orbital period of the CVs. 

The results for the identified CVs selected from our X-ray lists are summarized in the third to sixth lines in Table~\ref{table:tess}. For example, our eROSITA source list contains 173 identified CVs and confirm the orbital period from $N_{P_0}=30$ sources or the strongest signal at harmonics from $N_{P_0}=12$ sources. We also find possible signal of the orbital variation from $N_{non}=35$ sources. We can see that the distribution patterns of the sources in Table~\ref{table:tess} are roughly consistent across the five source lists. Although the direct application of the distribution to the CV-candidates may  be crude,  we may  anticipate that if the periodic signal are confirmed in the TESS data, the most likely explanation is the orbital period, with the first harmonic being a secondary possibility.

Figure~\ref{fig:period-mg} shows the observed period for the CVs and CV-candidates selected by cross-matching  with the eROSITA sources. We find in the left panel of the figure that the CVs of our list typically exhibit a period between 0.1~{\rm day} below the so-called period gap of the CVs \citep{1983A&A...124..267S,2018ApJ...868...60G} and 0.05~{\rm day}, which is known as the minimum orbital period in the standard binary evolution of CVs \citep{2010MmSAI..81..849R}. These observed periods are typical values for non-magnetic CVs or polar \citep[see][figure 18]{2022ApJ...936..134T}. As the right panel of Fig.~\ref{fig:period-mg} shows,  the CV-candidates of our list also have a similar distribution of the period of the identified CVs. Since our CV-candidates are selected from the bridge region between the WD cooling sequence and the main-sequence in H-R diagram, it is more likely that the origin of their periodic signal is related to the orbital period of accreting WDs candidates systems.

\begin{figure*}
\centering
\includegraphics[scale=0.45]{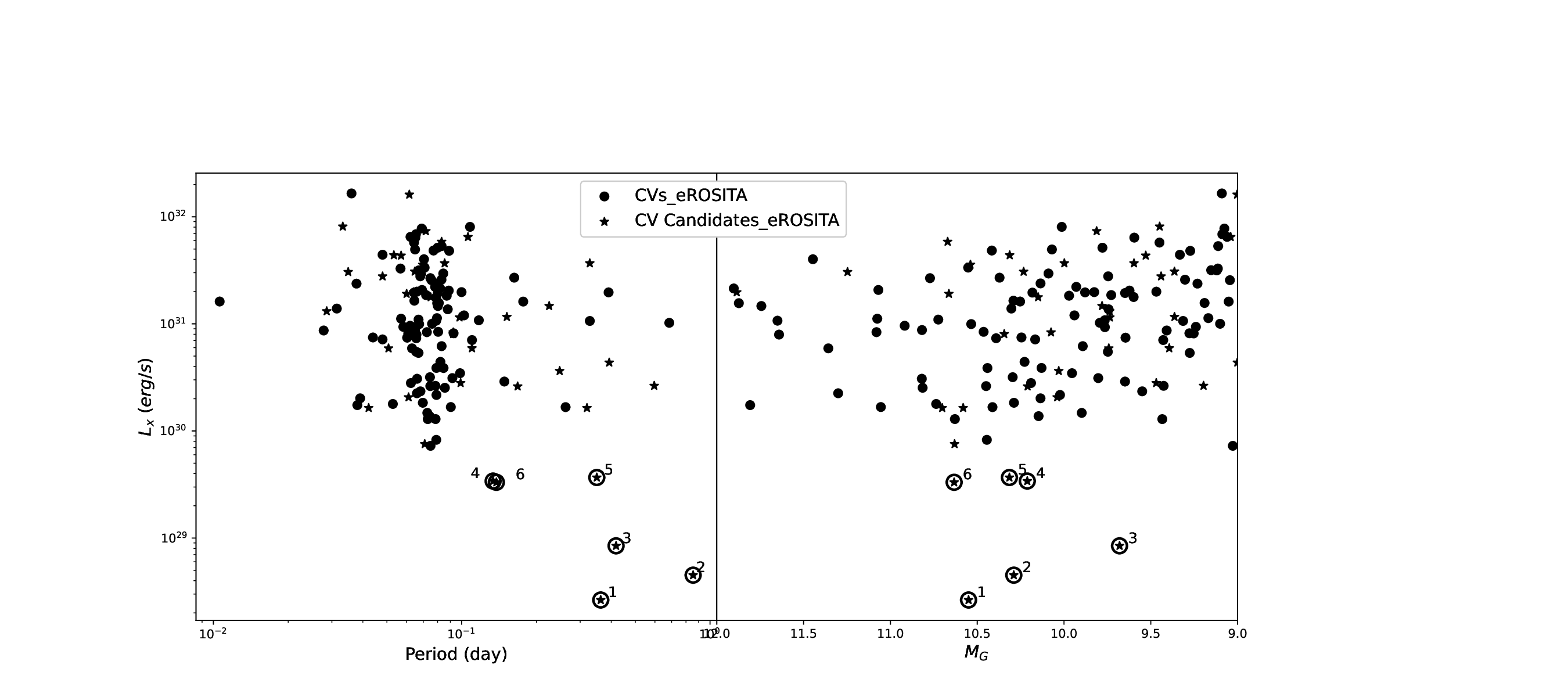}
\caption{ Left: Period and estimated X-ray luminosity in the eROSITA energy bands. Right: G-band absolute magnitude and the estimated X-ray luminosity. The symbols of the circles and stars corresponds to the CVs and CV-candidates respectively. The six  sources with a lower X-ray luminosity, indicated by the numbering, are presented in Table~\ref{table:six low luminosity sources}.} 
\label{period,Mg-Lx,diagram}
\centering
\end{figure*}

\subsection{Optical and X-ray properties}
\label{sec:xpro}
\begin{table*}
\centering
\caption{List of six source with low X-ray luminosity}
\begin{tabular}{cccccc}
\toprule
Number & GAIA ID & Name$^{a}$ & Luminosity & Period  & Type$^{b}$ \\

&&& ($10^{30}$ erg/s)& (days)  \\
\hline
1 & 4761389616087396096 & EC 05089-5933 & 0.0264 & 0.363(3) & WD \\
2 & 6576232101460909056 & HE 2123-4446 & 0.0451 & 0.86(2) & WD \\
3 & 5293314439454159488 &  & 0.0847 & 0.209(3)$^{c}$ & Eclipsing \\
4 & 6088378414265678336 & & 0.3406 & 0.1338(4) & \\
5 & 6286158119187026688 & HE 1432-1625 & 0.3678 & 0.35015(5)$^{d}$ & WD \\
6 & 5223042417940868864 & & 0.3317 & 0.1379(4) & \\
\hline
\end{tabular}
\begin{tablenotes}
\item{a}: Name listed in SIMBAD and/or VSX. 
\item{b}: Type recorded in SIMBAD and/or VSX.  
\item{c}: Period signal in TESS observation is the first harmony of the VSX recorded period.  
\item{d}: Period signal reported in ZTF observation.  
\end{tablenotes}
\label{table:six low luminosity sources}
\end{table*}

The eROSITA catalog provides the observed X-ray flux in five distinct energy bands:
$P_1=0.2-0.5$~keV, $P_2=0.5-1.0~$keV, $P_3=1.0-2.0$~keV, $P_4=2.0-5.0$~keV and $P_5=5.0-8.0$~keV energy bands. Using the information of the fluxes of eROSITA and the parallax of the corresponding  GAIA sources,  we estimate the luminosity in 0.2-8.0~keV  bands by assuming an isotropic radiation. Figure~\ref{period,Mg-Lx,diagram} plots the estimated X-ray luminosity with the observed period (left panel) or the GAIA G-band magnitudes (right panel) for the selected eROSITA soruces with a detection of the peridoic modulation in ZTF/TESS data:  the symbols marked by the circle and star correspond to  the identified  CVs and  CV-candidates, respectively.  We can see in the figure that majority of the sources have a luminosity in the range of $L_{X}\sim 10^{30-32}~{\rm erg~s^{-1}}$,  which is the characteristic of typical  X-ray emitting CVs.

\begin{figure*}
\centering
\includegraphics[scale=0.5]{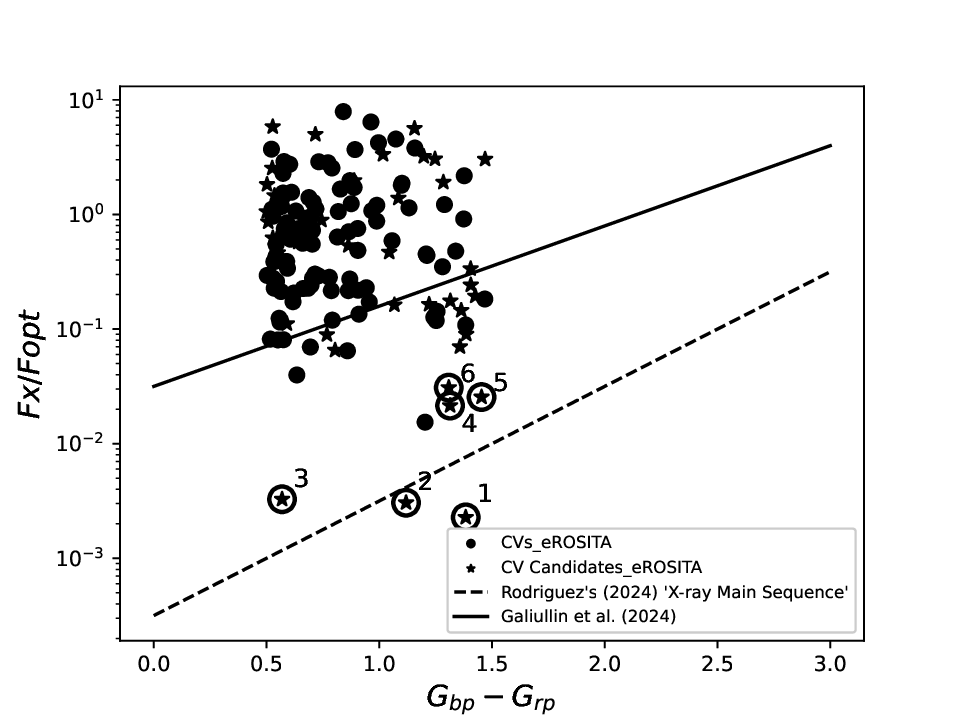}
\includegraphics[scale=0.4]{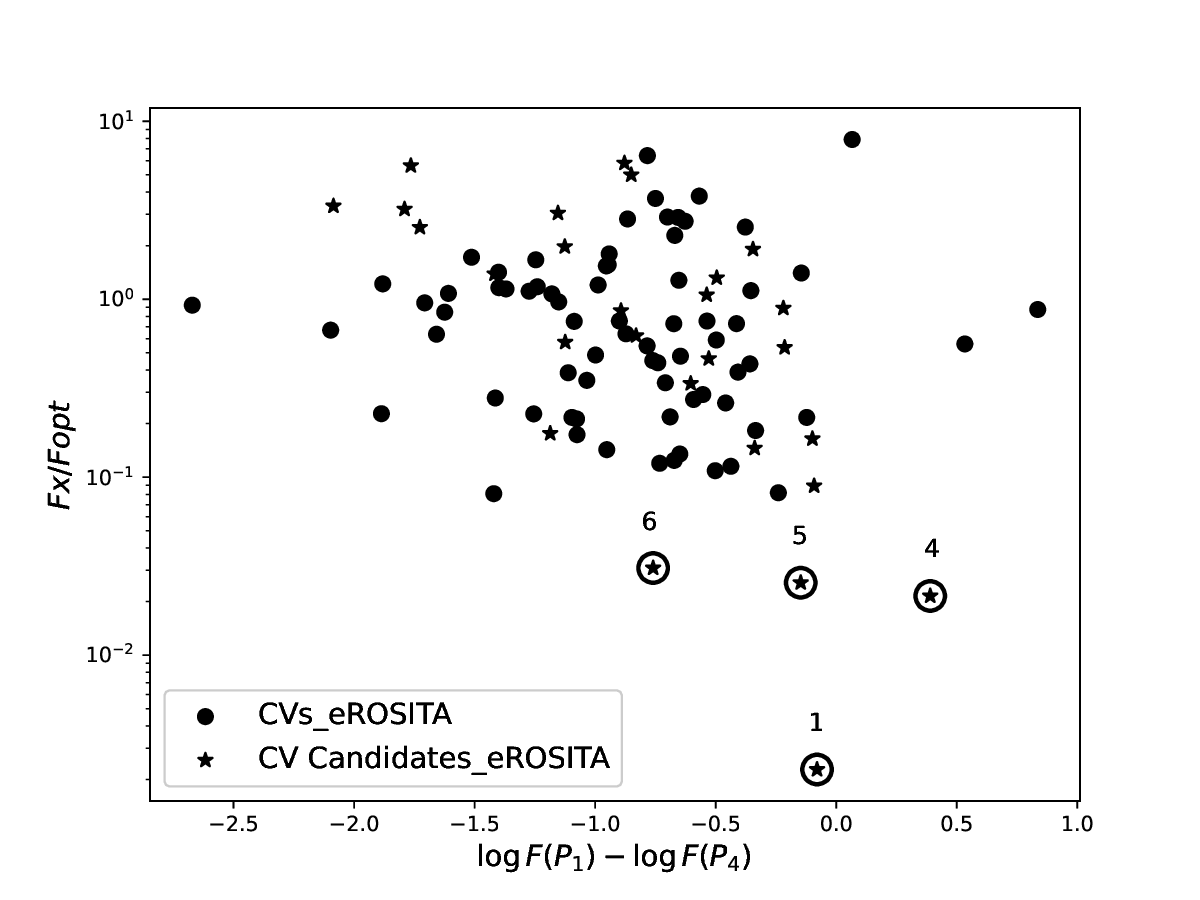}
\caption{Ratio of the X-ray flux ($F_X$) in the 0.2-8.0~keV energy bands to the optical flux ($F_{opt}$) vs. the hardness of the optical bands 
(left panel) or of the X-ray bands (right panel). The $F(P_1)$ and $F(P_4)$ represent the observed fluxes in 
$P_1=0.2-0.5$~keV bands and $P_4=2.0-5.0$~keV bands respectively. The circles and  stars corresponds to the identified CVs and CV-candidates for eROSITA sources with the detection of the periodic signal in ZTF/TESS data. The solid and dashed lines in the left pane is taken from \cite{2024A&A...690A.374G} and \cite{2024PASP..136e4201R}, respectively. The six sources enclosed by the black circle are low-luminosity candidates correspond to those in Fig. \ref{period,Mg-Lx,diagram} and Table~\ref{table:six low luminosity sources}. The second source is missing in the right panel as its $F(P_4)$ is not measured, while the third source is missing because neither  $F(P_1)$ nor $F(P_4)$ has been measured.} 
\label{fig:GAIAbprp-flux}
\end{figure*}

Figure~\ref{period,Mg-Lx,diagram} also shows six sources, indicated by the numbering, which exhibit the X-ray luminosity smaller than  $L_{X}\sim 10^{30}~{\rm erg~s^{-1}}$. To investigate the possible contamination from the other X-ray sources,  Fig.~\ref{fig:GAIAbprp-flux} presents  X-ray-to-optical flux ratio ($F_x/F_{opt}$) as a function of the optical color (left panel)  or the X-ray color (right panel) for the eROSITA sources with the detection of the periodic modulation. \cite{2024PASP..136e4201R} proposes an empirical boundary to discriminate the accreting compact binary systems from the X-ray emitting active stars. We find that those six sources with lower X-ray luminosity are clustered around the boundary, which is the dashed line in Fig.~\ref{fig:GAIAbprp-flux}. The light curve profiles with the ZTF/TESS data and the characteristics of those six sources are presented in Sect.~\ref{sec:lowx} and Table~\ref{table:six low luminosity sources}, respectively.  We find in Table~\ref{table:six low luminosity sources} that one source is an eclipsing binary in VSX and three sources are categorized as WD in SIMBAD. Thus, the most probable explanation of those six sources is the eclipsing binary with the companion star's activity likely producing the observed X-ray emission. Nevertheless, since other X-ray emission processes, such as low accretion rate, cannot be entirely ruled out in this study, the deeper observations would be needed to classify those sources conclusively.

\begin{figure}
\includegraphics[scale=0.5]{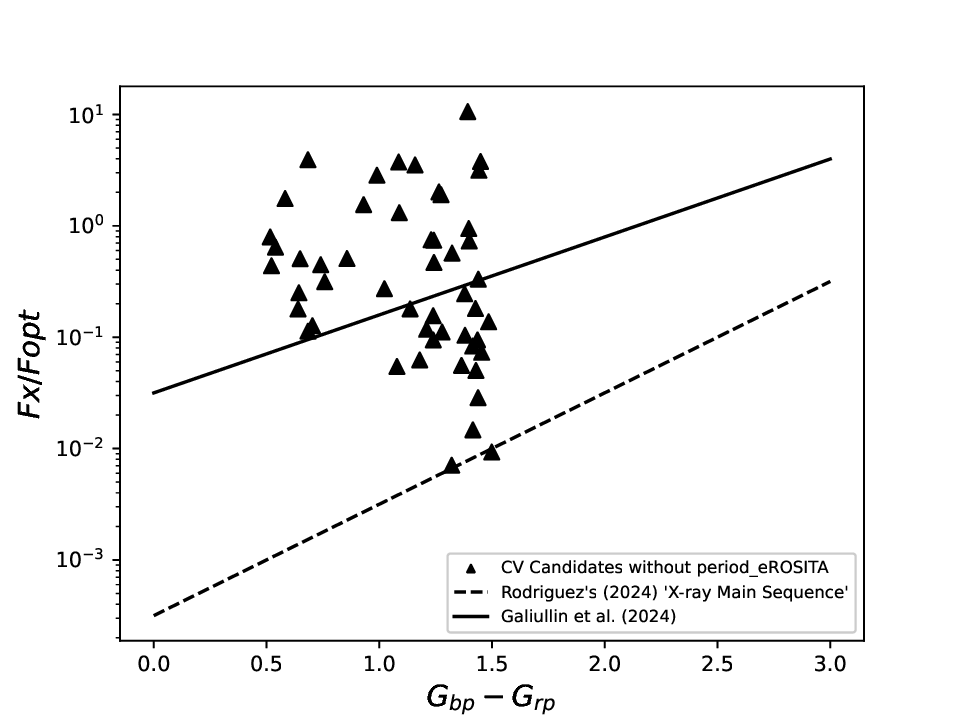}
\caption{The same as for Fig.~\ref{fig:GAIAbprp-flux}, but the CV-candidates are without detection of the periodic signal in ZTF/TESS data. }
\label{fig:GAIAbprp-flux1}
\end{figure}

\cite{2024A&A...690A.374G} also proposes a boundary (solid line in the left panel of Fig.~\ref{fig:GAIAbprp-flux}) to select pure sample of CVs. We can see that most of our CV-candidates are located around or above this boundary, consistent with the properties of the identified CVs. There are 24 CV-candidates are located above this boundary, suggesting their periodic modulation in ZTF/TESS data 
are probably related to the orbital periods. For the CV-candidates without periodic signals, Fig.~\ref{fig:GAIAbprp-flux1} presents their flux ratio ($F_{\rm X}/F_{\rm opt}$)  and the optical color ($G_{\rm bp}-G_{\rm rp}$). We find  31 sources located above the boundary to select pure sample of CV-candidates.Their locations suggest they are promising CV-candidates, but further observations are needed to confirm their nature due to the lack of periodic signals. 

\subsection{Differentiating from other compact object binary systems}
\label{sec:dif}
Although the detection of the periodic signal and estimated 
X-ray luminosity of most of the CV-candidates are consistent to  those of known CVs, the current photometric study may still occur a contamination 
from other type of compact object binary systems, such as Galactic nucleus (AGNs) and high-mass/low-mass X-ray binaries. 
As pointed out by \cite{2024A&A...686A.110S}, the CVs and AGNs show similar  optical and X-ray flux ratios and may be difficult to discriminates between two candidates based on the color-color diagram.   We, therefore, perform cross-matching our list of the eROSITA sources with GAIA DR3 AGN catalog \footnote{\url{https://vizier.cds.unistra.fr/viz-bin/VizieR-3?-source=I/358/vagn}} \citep{2022yCat.1358....0G} and obtain no matches. Consequently, AGN contamination would be minimal.
To effectively distinguish between these two types of the X-ray sources,  a detailed analysis of the X-ray spectral  properties   will be necessary.  Additionally, the investigation of the variability patterns could discriminate between the CVs and AGNs: CVs have a short-term periodic variations with orbital motion  typically less
than half a day, while AGNs usually show the long-term variability from days to years~\citep{1997ARA&A..35..445U,2006ConPh..47..363S}.

\begin{figure}
\includegraphics[scale=0.5]{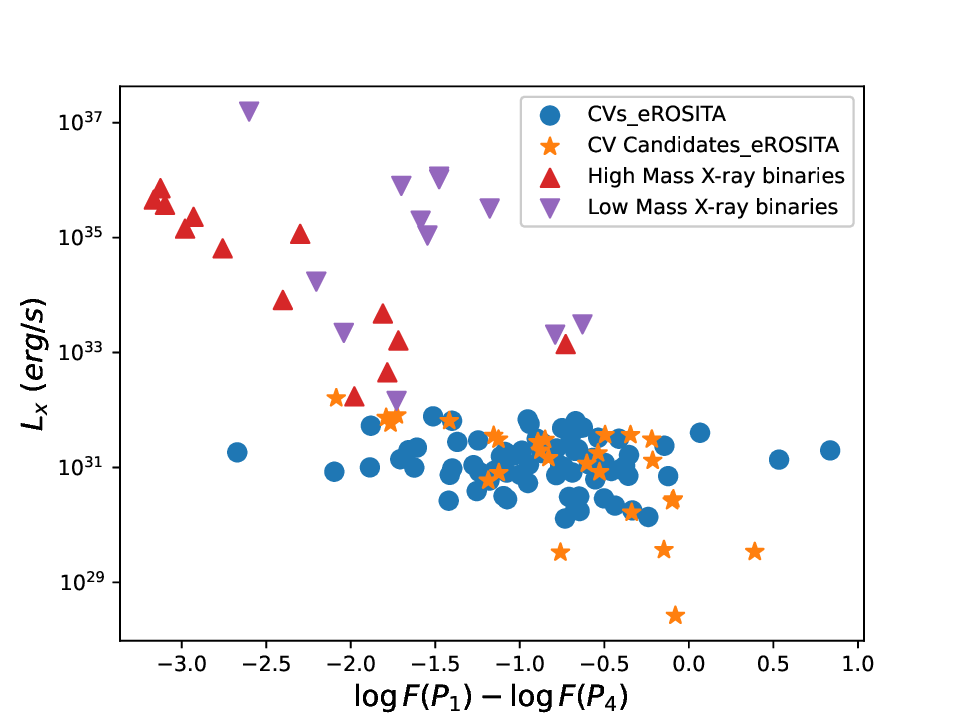}
\caption{Luminosity and hardness relation measured by eROSITA. The triangles and inverted triangles correspond to the high-mass and low-mass X-ray binaries, respectively. The circles, stars correspond to the CVs, CV-candidates of eROSITA. }
\label{fig:P1P4-Lx}
\end{figure}

The Galactic high-mass and low-mass X-ray binaries hosting a neutron star or black hole may be another main source of the contamination in the X-ray bands. In the optical bands, however, the CVs usually have a GAIA magnitude smaller than that of the high-mass binary system.  Moreover, the X-ray binaries usually show a  higher X-ray luminosity ($>10^{33}~{\rm erg~s^{-1}}$)  than that of the CVs. Figure~\ref{fig:P1P4-Lx}  illustrates the  difference of the X-ray properties of the eROSITA observations  among three types of the binary systems; we cross-match the eROSITA catalog with 
X-ray binary catalogs\footnote{\url{https://binary-revolution.github.io/HMXBwebcat/}} \footnote{\url{https://binary-revolution.github.io/LMXBwebcat}} \citep{2023A&A...671A.149F,2024A&A...684A.124F}, using the limit of the angular separation of 10". 
As shown in the figure, the high-mass and low-mass X-ray binaries generally exhibit higher X-ray luminosity ($>10^{33}~{\rm erg~s^{-1}}$) compared to CVs. Additionally,  the hardness ratio (${\rm log}F(P_1)-{\rm log}F(P_4)$) increases with decreasing of the luminosity in the eROSITA bands. For CVs, on the other hand, the hardness ratio is less dependent on the luminosity. 
The figure indicates that our eROSITA sources identified as CV-candidates are more consistent with the CVs rather than high-mass or low-mass X-ray binaries.

\begin{figure}
\includegraphics[scale=0.5]{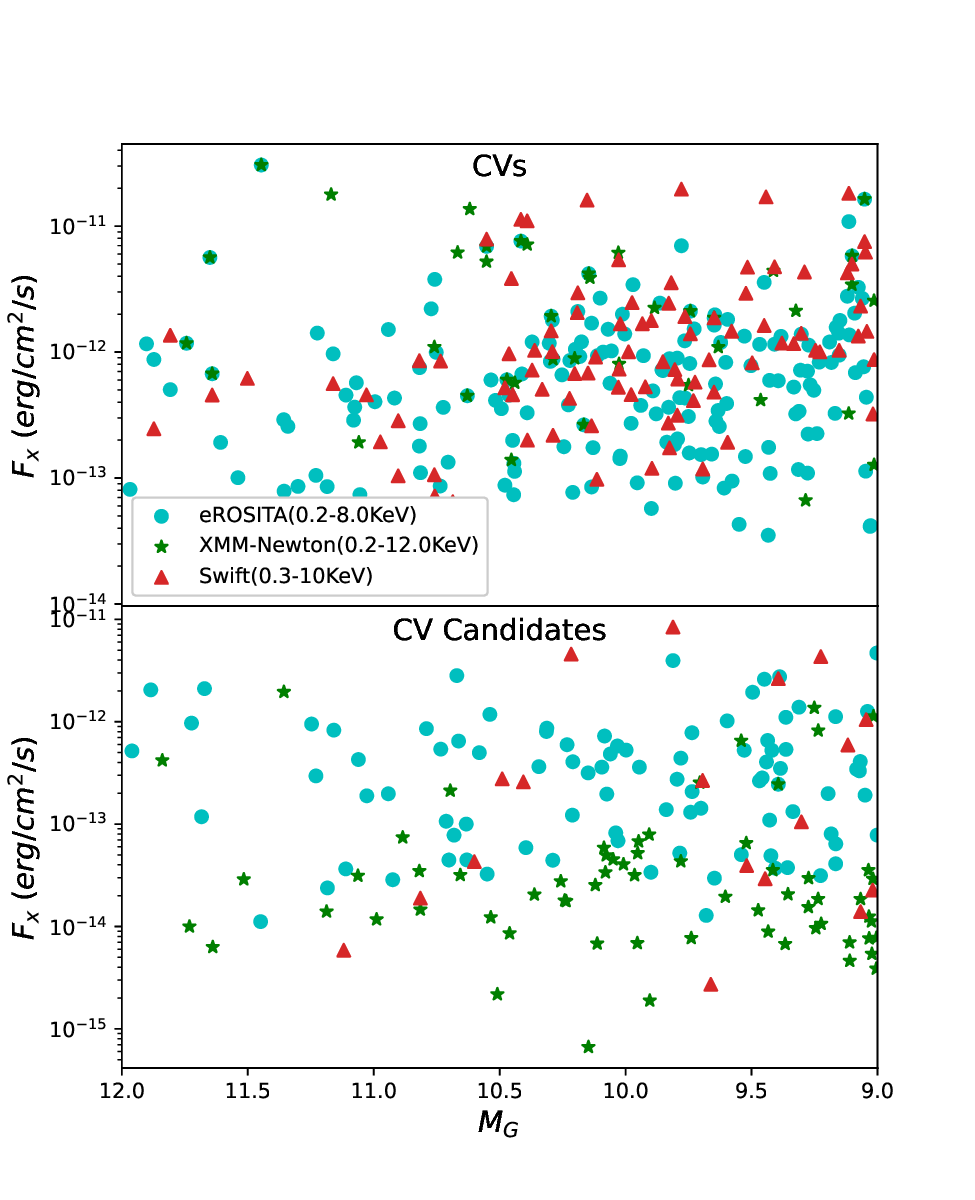}
\caption{Flux distribution of CVs (upper panel) and CV-candidates (bottom panel) in eROSITA (filled circles), XMM-Newton (stars) and Swift (triangles).}
\label{fig:Fx_CV}
\end{figure}

\begin{figure}
\includegraphics[scale=0.5]{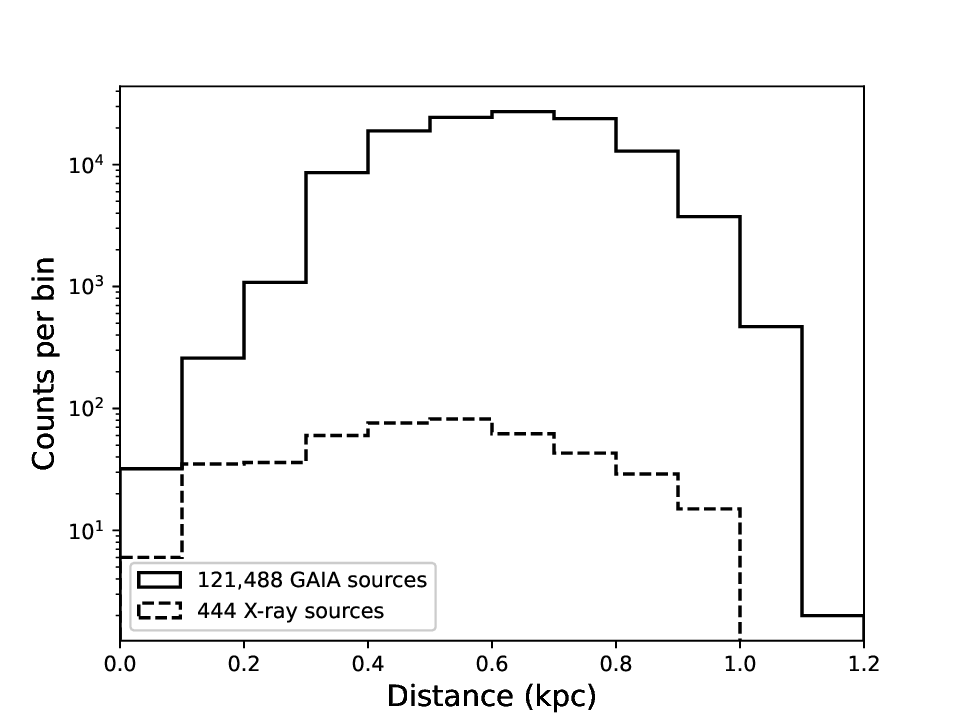}
\caption{Distribution of the source distances for 121,488 GAIA sources (solid histogram) selected in section~\ref{sec:gdata} and for 444  GAIA sources that  may be associated with X-ray sources (dashed histogram). }
\label{fig:dist}
\end{figure}

\subsection{Comparison among X-ray catalogs}
\label{sec:comp}
As expected and demonstrated in Table~1, the eROSITA can observe the X-ray emission from more identified CVs and  CV-candidates  compared to other X-ray observations.  To illustrate  the flux sensitivity of each catalog,  Fig.~\ref{fig:Fx_CV} presents the observed X-ray fluxes of the identified CVs (upper panel) and CV-candidates (lower panel) selected from the eROSITA (circles), XMM-Newton (stars) and Swift (triangles). We remove the ROSAT sources as (i) most CVs selected from ROSAT can be found in other catalogs, and (ii) number of CV-candidates is much less than those of other catalogs. The three catalogs provide the energy flux  in different energy bands: 0.2-8.0~keV bands of eROSITA, 0.2-12.0~keV bands of XMM-Newton and 0.3-10~keV bands of Swift. Despite these differences, those fluxes will represent the order of magnitude of the soft X-ray emission from the system.

As illustrated in upper panel of  Fig.~\ref{fig:Fx_CV},  the typical fluxes of identified CVs are of the order of $10^{-14}-10^{-11}~{\rm erg~cm^{-2}s^{-1}}$, and the distributions of the three observations are similar to each other. Notably,  about 120 CVs from the eROSITA catalog are not found in other three catalogs.  This demonstrates that the eROSITA observation is more efficient to discover the X-ray emission from the CVs and encourages to carry out a target observation for further in-depth study~\citep[see also][]{2024A&A...690A.243S}. Despite the eROSITA mainly contributes to our CV-candidates, it does not cover about 64 (out of 65) and 18 (out of 20) CV-candidates selected from  the XMM-Newton and Swift catalogs, respectively. This is mainly because  those sources are located at the Galactic hemisphere that is not covered by the current eROSITA catalog.

The CV-candidates in the bottom panel of Fig.~\ref{fig:Fx_CV} contains 91 sources from the eROSITA, 65 from the XMM-Newton and 20 from Swift.  Out of the 65 XMM-Newton sources and 20 Swift sources, only one XMM-Newton source and two Swift sources are found in the eROSITA catalog, and about half sources are located at the Galactic hemisphere that is not covered by the current eROSITA catalog.  We find in the figure that the flux distributions of the eROSITA sources and the Swift sources are  similar to that for the CV ($F_X\sim 10^{-14}-10^{-11}~{\rm erg~cm^{-2}s^{-1}}$).  For XMM-Newton sources (the symbol with stars), on the other hand, a certain fraction of the sources have a flux below $10^{-14}~{\rm erg~cm^{-2}s^{-1}}$.   From Fig.~\ref{fig:Fx_CV}, we may draw two conclusions.  First,  the 
large population of X-ray emitting CV-candidates having a flux of $F_X\sim 10^{-14}-10^{-11}~{\rm erg~cm^{-2}s^{-1}}$ have newly confirmed by  the eROSITA  due to its larger covering sky region.  Second, the XMM-Newton observations indicate a significant population of the  CV-candidates having a flux  below  $\sim 10^{14}~{\rm erg~cm^{-2}s^{-1}}$. So eROSITA observation with more exposure time has a potential   to find more X-ray emitting CV-candidates with a flux below $\sim 10^{14}~{\rm erg~cm^{-2}s^{-1}}$.  Since the current eROSITA catalog is result of the operation in the first six month, it will be expected that the future operation will further  increase  the population of the X-ray identified  CVs and CV-candidates.

\subsection{X-ray emitting, accreting WD binaries within 100pc}

\begin{table*}
\caption{List of X-ray emitting, accreting WD binaries within 100~pc in this study}
\label{table:xray}
\centering           
\begin{tabular}{llllll}
\toprule
Name & Distance & Period & Luminosity $^{a}$  & Catalogs & Category$^{b}$\\
& (pc) & (days) &  ($\rm 10^{30}~{\rm erg~s^{-1}}$) &\\
\hline
U Gem & 93 & 0.177(3) & 16 & eROSITA, XMM, Swift, ROSAT &CV \\
VW Hyi & 54 & 0.0761(4) & 1.4& eROSITA, XMM, ROSAT &CV \\
HE 1432-1625  & 64 & 0.35015(5) & 0.37 & eROSITA &WD \\
HE 2123-4446  &  95 & 0.86(2)  & 0.04 & eROSITA &WD \\
EC 05089-5933  & 84 & 0.363(3) & 0.02 & eROSITA  &WD\\
UZ Sex & 40&  0.60(2)  & 0.01  & XMM&WD\\
\hline
AR Uma$^{c}$  & 99 &0.0805(4) & 400 & & CV \\
\hline
\end{tabular}
\begin{tablenotes}
\item{$a$}: Luminosity in 0.2-8.0~keV bands for the five eROSITA sources and in 0.2-12~keV bands for UZ Sex. 
\item{$b$}: SIMBAD source category.
\item{$c$}: The luminous X-ray CV that cannot be found in the four catalogs. The luminosity in 0.1-3.0~keV is referred from \cite{1994ApJ...426..288R}.
\end{tablenotes}
\end{table*}
A significant fraction of accreting WD binaries are expected to be X-ray emitters due to either (i) the accretion process, as seen in CVs, or (ii) hydrogen or helium burning on the surface, as in supersoft sources with a luminosity of $10^{37-38}~{\rm erg~s^{-1}}$.  The new eROSITA all-sky survey presents a valuable opportunity to explore Galactic distribution  of the accreting WDs candidates. Figure~\ref{fig:dist} show the distribution of the distance for  121,488 GAIA sources  (solid line) and 444 X-ray sources (dashed line), respectively.

Due to the high completeness of the accreting WD samples within 100~pc by the GAIA  observations \citep{2024ApJ...970L..11H},  we investigate our targets located within 100~pc.  Among 121,488 GAIA sources in our study,  32 sources are located within 100 pc and they are listed in Table~\ref{table:100pc}. We find that 7 sources among them exhibit  possible X-ray counterparts and summarize their characteristics in  Table~\ref{table:xray}. 
Three systems, U~Gem, VW~Hyi and AR Uma, are identified CVs. 
The remaining four sources  exhibit a X-ray luminosity less than $10^{30}~{\rm erg~s^{-1}}$, and  three of them have already mentioned as a low luminous X-ray sources in Table~\ref{table:six low luminosity sources}.

As shown in Table~\ref{table:xray}, there are two  X-ray emitting, accreting WD binaries, but there are no record in the eROSITA catalog. 
First, UZ~Sex is the closest binary system in our targets, but its X-ray flux with $\sim 10^{-14}~{\rm erg~cm^{-2}~s^{-1}}$ measured by the XMM-Newton is the lowest, as Table~4 indicates. Hence, the current eROSITA catalog does not include UZ~SeX, likely due to the limit of the  sensitivity of the current survey. Second, AR UMa, which is an identified CV, has a short orbital period with $\sim 0.0805(4)$~day and it has not been recorded in the four X-ray catalogs used in this study. However, \cite{1994ApJ...426..288R} report AR~UMa as an X-ray source with a luminosity of $\sim  4\times 10^{32}~{\rm erg~s^{-1}}$, and the source has been  recorded in the ROSAT Source Catalog of Pointed Observations with the High Resolution Imager~\citep{2000yCat.9028....0R}. In addition, recent  target observation  of AR UMa by XMM-Newton  (Obs.IDn: 0884870101 and  PI: Schwope, Axel) clearly shows  an bright X-ray emission that modulates with the orbital period, suggesting a polar.  Unfortunately,  the current eROSITA catalog, which records the X-ray sources located at western Galactic hemisphere \citep{2024A&A...682A..34M}, 
does not cover the sky region around the source. 

Among 32 sources located within 100~pc, we confirm the X-ray emissions from 7 sources. In Table~\ref{table:100pc}, on the other hand, there are  several sources (e.g., GD~245, UCAC4 135-002106, etc.) that have an orbital period shorter than one day,  but no X-ray emissions have been recorded in the catalogs. This maybe because the sources have not been covered by the previous observations. Other possibility is that the source is an eclipsing binary system rather than CV and the binary system has a weak or non accretion process.  For cases such as UCAC4 135-002106  and UCAC4 293-078484, for example, the eROSITA catalog converse the source regions, suggesting their X-ray fluxes are of the order of or below  $\sim 10^{-14}~{\rm erg~cm^{-2}s^{-1}}$. For other cases, such as GD 245, the current eROSITA catalog does not cover their sky regions. Hence, the results of the continuous operation and source list of the all sky of the eROSITA observation  are desired to obtain more complete list of the X-ray emitting, accreting WD binaries within 100~pc. 

\section{Summary}
The sky survey of eROSITA offers a new opportunity to identify for numerous  X-ray emitting CVs. In this study, therefore,  we have searched for candidates of the X-ray emitting CVs by cross-matching between four X-ray catalogs (eROSITA, XMM-Newton, Swift and ROSAT ) and the GAIA's sources. We have demonstrated how the eROSITA survey is more efficient in searching for X-ray emitting CVs.  We have selected 264 sources,  including 176 identified CVs and 91 CV-candidates, from the eROSITA catalog. Among 91  CV-candidates, we have identified a periodic signal from 40 sources in ZTF and/or TESS photometric light curves. The distribution of the period and the expected X-ray luminosity are consistent with those of the identified CVs (Figs.~\ref{fig:period-mg} and~\ref{period,Mg-Lx,diagram}).  

 Most of our CV-candidates with the detection of the periodic modulation are discriminated from the active starts on the color-color diagram using the X-ray and optical observations (Fig.~\ref{fig:GAIAbprp-flux}). On the other hand, we also find six sources whose X-ray luminosity is $L_X<10^{30}~{\rm erg~cm^{-2}s^{-1}}$, which is smaller than typical CVs (Fig.~\ref{period,Mg-Lx,diagram}). This suggests that those sources are candidates of the eclipsing binary rather than CVs (Table~\ref{table:six low luminosity sources}).  The X-ray emission from AGNs may be another source of the  contamination. Although a cross-matching between list of our eROSITA sources and AGN-candidates provides null result, further studies of the X-ray properties (e.g., spectrum and modulation) are necessary to differentiate from X-ray emitting CVs and AGNs. We also demonstrated that the CV-candidates are discriminated from the low-mass and high-mass X-ray binaries (Fig.~\ref{fig:P1P4-Lx} ).

Due to its larger covering sky region, the number of the CVs and CV-candidates selected from eROSITA catalogs (264 sources) are much more than those from XMM-Newton (109), Swift (111) and ROSAT (69), as presented in Table~\ref{table:cross}. It is found that while  the eROSITA sources selected in this study have a bigger flux than $10^{-14}~{\rm erg~cm^{-2}s^{-1}}$, the XMM-Newton observations indicates a certain population of the CV-candidates with a flux below $10^{-14}~{\rm erg~cm^{-2}s^{-1}}$ (Fig.~\ref{fig:Fx_CV}).  This suggests that the future operation of eROSITA observation will increase the population of the CVs and CV-candidates identified by the X-ray bands.  

Among 121,488 GAIA sources selected in this study, 32 sources are located within 100~pc and 7 sources exhibit X-ray emissions. Three sources are identified CVs and other four sources are likely eclipsing binary systems. The five sources can be found in the eROSITA catalog, and  three out of them can not be found in other three catalogs. On the other hand, current eROSITA catalog will miss some X-ray emitting, accreting WD binaries located within 100~pc that have a  flux smaller than the current sensitivity of the survey (e.g., UZ Sex). Moreover, the current eROSITA catalog covers only one Galactic hemisphere. Hence, the continuous eROSITA survey and  the catalog covering whole sky region will provide a more comprehensive understanding for the population of the X-ray emitting, accreting WD binaries.

\section{Data availability}
Tables of identified CVs and CV-candidates find in this study are only available in electronic form at the CDS via anonymous ftp to cdsarc.u-strasbg.fr (130.79.128.5) or via http://cdsweb.u-strasbg.fr/cgi-bin/qcat?J/A+A/. 

\begin{acknowledgements}
  We express our thanks to the referee for his/her comments and suggestions, which have significantly improved our manuscript. We appreciate Dr. A.D. Schwope for providing the list of CVs selected from the eROSITA catalog.  We are grateful to Drs A.K.H. Kong, J.Mao, X. Hou, K.K. Li, L.C.-C. Lin and K.L. Li for useful discussion for the CVs. 
  This research has made use of the SIMBAD database,operated at CDS, Strasbourg, France and the International Variable Star Index (VSX) database, operated at AAVSO, Cambridge, Massachusetts, USA.   This work is based on data from eROSITA, the soft X-ray instrument aboard SRG, a joint Russian-German science mission supported by the Russian Space Agency (Roskosmos), in the interests of the Russian Academy of Sciences represented by its Space Research Institute (IKI), and the Deutsches Zentrum für Luft- und Raumfahrt (DLR). The SRG spacecraft was built by Lavochkin Association (NPOL) and its subcontractors, and is operated by NPOL with support from the Max Planck Institute for Extraterrestrial Physics (MPE). The development and construction of the eROSITA X-ray instrument was led by MPE, with contributions from the Dr. Karl Remeis Observatory Bamberg \& ECAP (FAU Erlangen-Nuernberg), the University of Hamburg Observatory, the Leibniz Institute for Astrophysics Potsdam (AIP), and the Institute for Astronomy and Astrophysics of the University of Tübingen, with the support of DLR and the Max Planck Society. The Argelander Institute for Astronomy of the University of Bonn and the Ludwig Maximilians
  University  at  Munich also participated in the science preparation for eROSITA. X.X.W. and J.T. are supported by the National Key Research and Development Program of China (grant No. 2020YFC2201400) and the National Natural Science Foundation of China (grant No. 12173014).

\end{acknowledgements}

%
%

\bibliography{aanda.bib}
\bibliographystyle{aa}

\begin{appendix} 

\onecolumn

\section{Light curve of six sources with lower X-ray luminosity}
\label{sec:lowx}
\begin{figure}[htp]
    \centering
    \begin{minipage}{0.45\textwidth} 
        \centering
        \includegraphics[width=\linewidth]{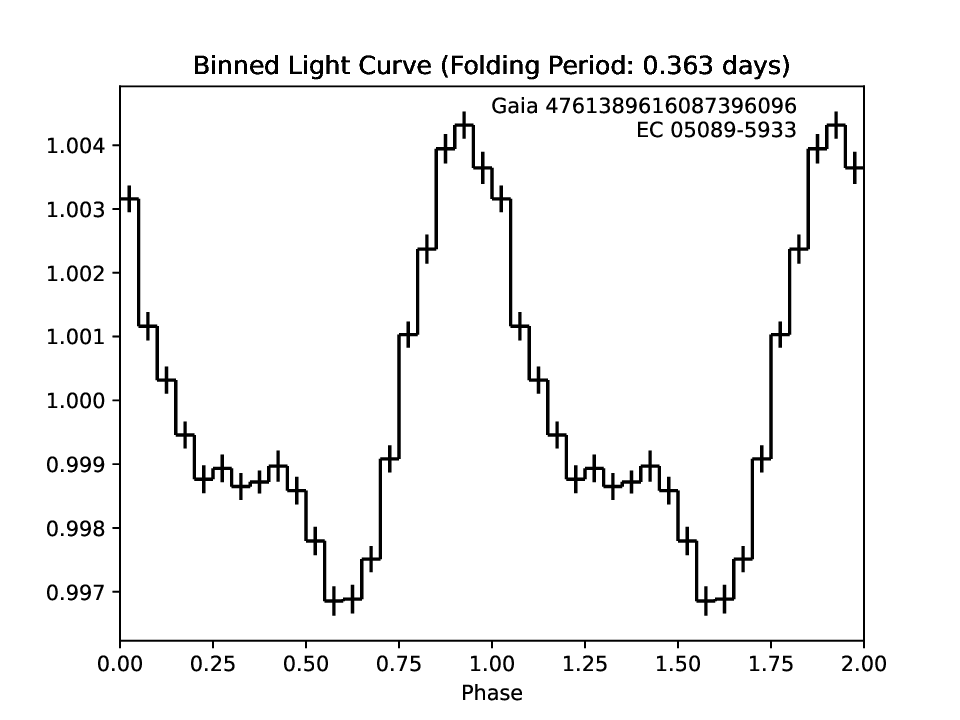}
        \subcaption{} \label{fig:subfig_a}
    \end{minipage}
    \hspace{0.05\textwidth} 
    \begin{minipage}{0.45\textwidth}  
        \centering
        \includegraphics[width=\linewidth]{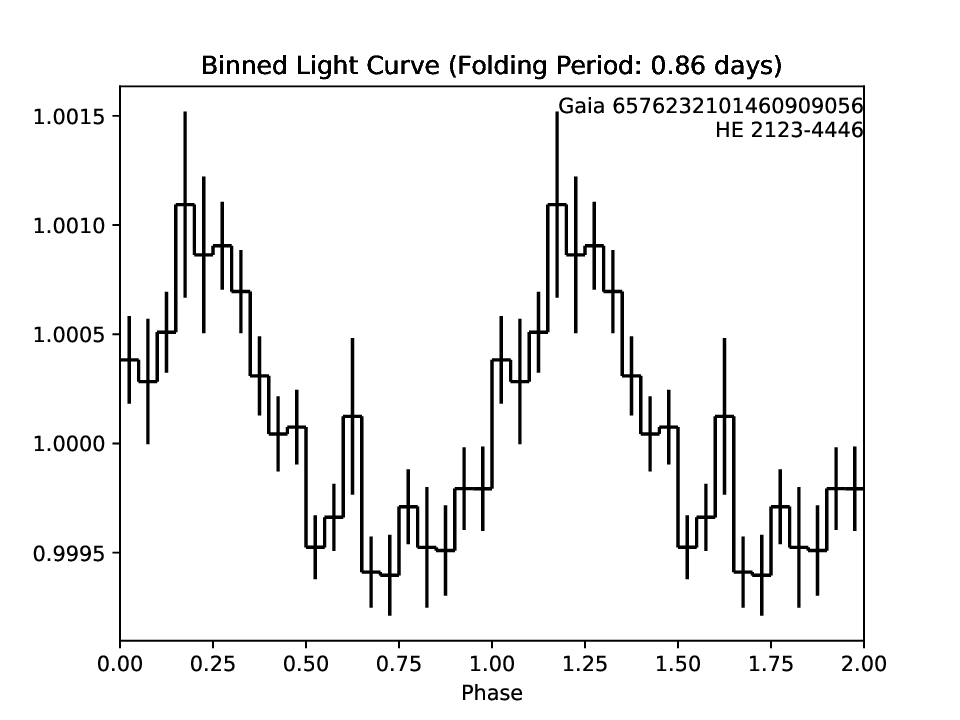}
        \subcaption{} \label{fig:subfig_b}
    \end{minipage}
    \vspace{0.1em}  
    \begin{minipage}{0.45\textwidth}
        \centering
        \includegraphics[width=\linewidth]{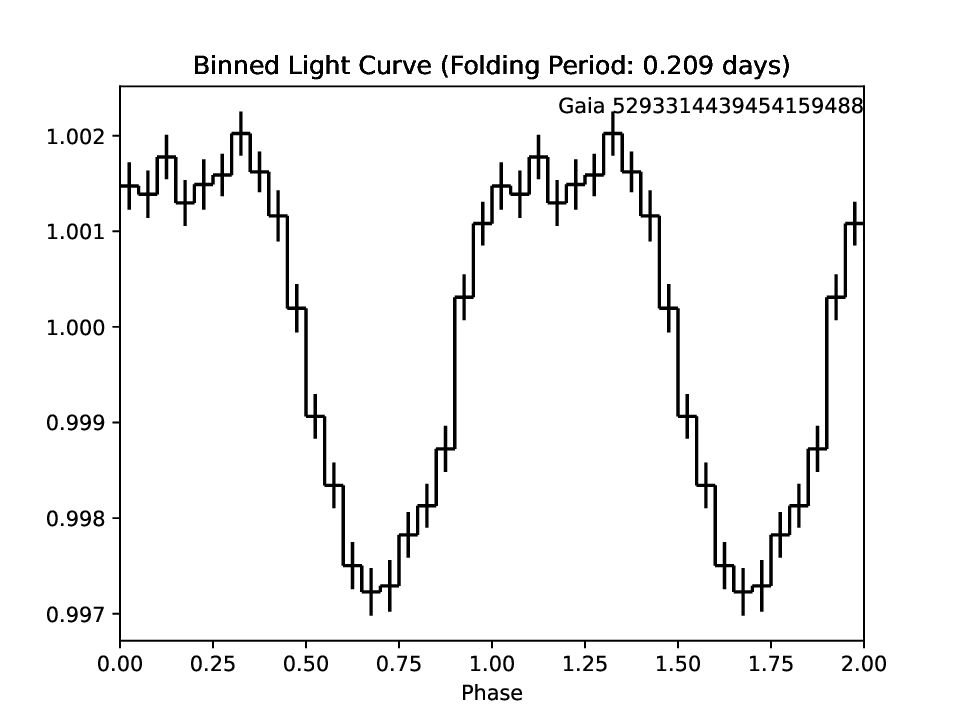}
        \subcaption{} \label{fig:subfig_c}
    \end{minipage}
    \hspace{0.05\textwidth}
    \begin{minipage}{0.45\textwidth}
        \centering
        \includegraphics[width=\linewidth]{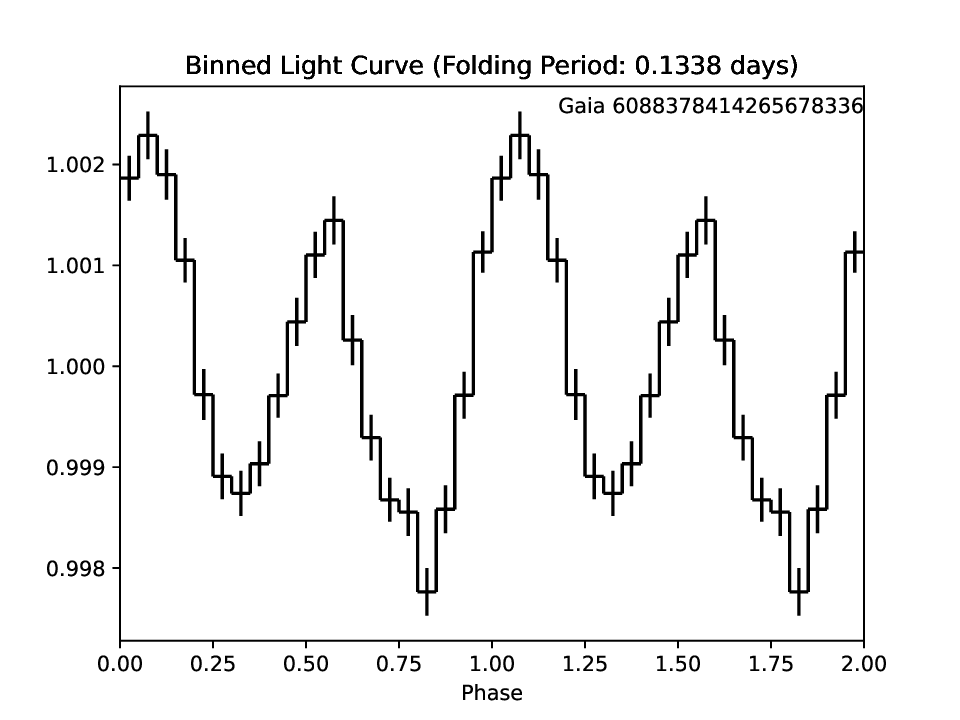}
        \subcaption{} \label{fig:subfig_d}
    \end{minipage}
    \vspace{0.1em}
    \begin{minipage}{0.45\textwidth}
        \centering
        \includegraphics[width=\linewidth]{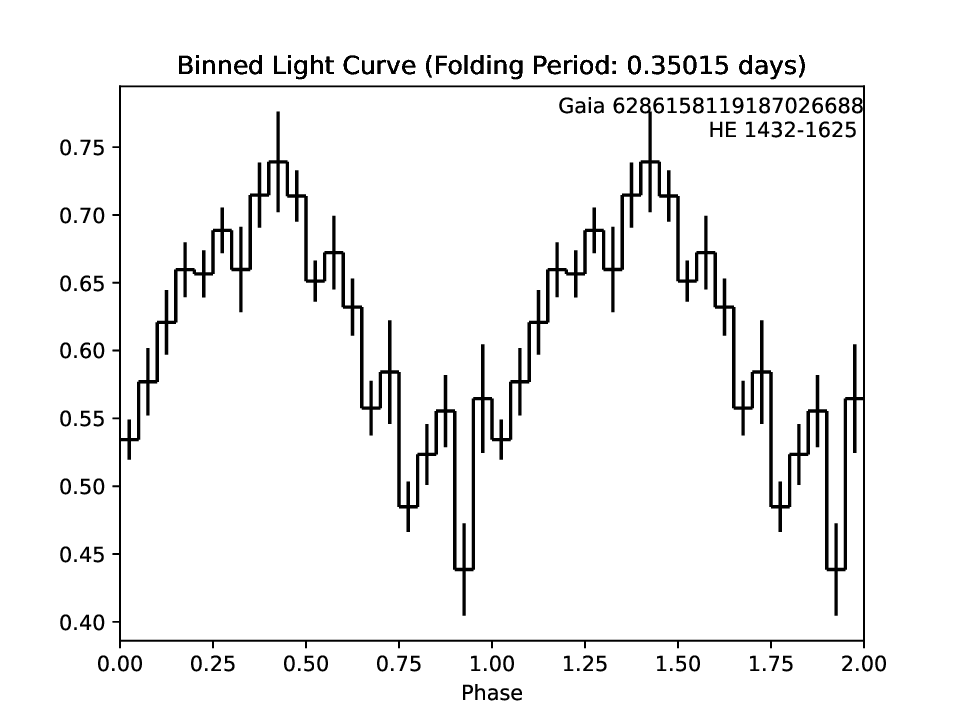}
        \subcaption{} \label{fig:subfig_e}
    \end{minipage}
    \hspace{0.05\textwidth}
    \begin{minipage}{0.45\textwidth}
        \centering
        \includegraphics[width=\linewidth]{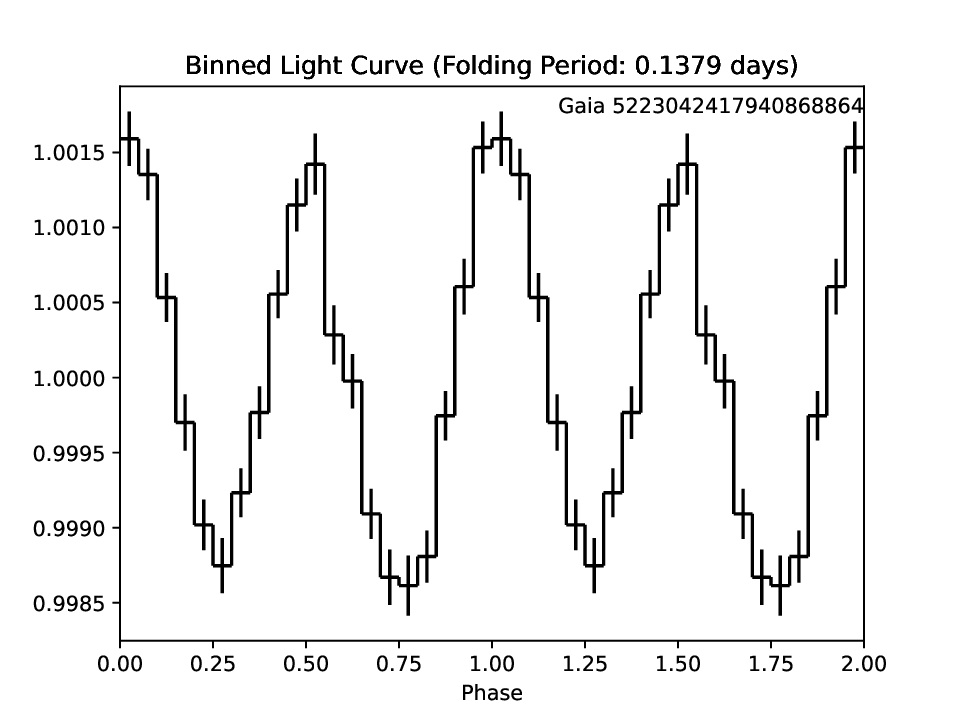}
        \subcaption{} \label{fig:subfig_f}
    \end{minipage}
    \caption{Phase-folding light curve for six low luminosity sources}
    \label{fig:six low luminosity sources}
\end{figure}

Figure~\ref{fig:six low luminosity sources} present the TESS light curves of the six CV-candidates with lower X-ray luminosity ($L_X<10^{30}~{\rm erg~s^{-1}}$, as present in Fig.~\ref{fig:GAIAbprp-flux} and Table~\ref{table:six low luminosity sources}.

\section{List of sources within 100~pc}
Table~\ref{table:100pc} provides the list of the selected GAIA sources located within 100~pc, based on their parallax. The fifth column shows the name of the WD or CVs indicated in the SIMBAD and/or VSX databases.  The sixth column summarizes the observed period of the modulation in the TESS light curves or the period reported in the VSX database.

\begin{table*}[ht]
\caption{List of selected GAIA sources within 100~pc}
\label{table:100pc}
\centering           
\begin{tabular}{llllll}
\toprule
GAIA ID & R.A. & Dec. & Distance  & Name$^{a}$  & Period\\
& (degree) & (degree) &  (pc) &  & (days) \\
\hline
1015082573717077376 & 131.45 & 48.88 & 70.2 & &  \\
150804964714568320 &       64.22 &       26.41&      64.3  & LAMOST J041653.27+262418.8 &  0.0059 and 1.07\\   
 1600448820449332864&      227.23 &      55.64&     76.1   &  & 10.32\\\
 1816934070031522688 &     306.13 &       20.48&     70.1& & $0.202$ \\  
 2791988540677395712  &    16.63  &      23.16 &     82.2   & & \\
 2843374388402149632   &   344.70 &       25.26&        50.1 & GD 245&  0.174$^{b}$\\ 
 2864507860881192320 &     354.80 &       25.87&      89.7   &  & \\
 3387904051723656320 &     77.03  &      11.19 &       92.3  & UCAC4 506-010126& 4.3 \\
 3547632785950758272 &     176.11 &     -16.25 &       70.4  &  & 0.554\\
 3777028806000016896 &     162.69 &      -4.74 &      99.3  &   & \\
 379359102453525376  &     6.64   &     39.15  &     63.5  &  G 171-B10A& \\
 3831059120921201280 &     157.15 &      -0.0085&   40.0  & {\bf UZ Sex}& 0.597$^{b}$\\
 394161419479425152  &     5.00   &     49.07   &     87.9 & &\\
 3969789309067081728 &     165.91 &       15.92 &       99.6 &   PM J11036+1555B & \\
 4263034940037993472 &     288.47 &      -1.09  &     77.0  &  &\\
 4333046892662353152 &     254.40 &      -12.94 &       90.6 &   &  \\
 4411741894799595776 &     242.23 &       1.76  &     44.3  &  HS 1606+0153&  \\
 4437836226304032384 &     244.29 &       5.51  &     69.4 &  PM J16171+0530 & \\
 4653893040002306432 &     62.30  &     -71.29  &     53.8 &   {\bf VW Hyi} & 0.0742$^{b}$\\
 4701214616008598144 &     34.80  &     -63.12  &     78.6 &  UCAC4 135-002106&  0.11 \\
 4761389616087396096 &     77.43  &     -59.49  &    84.7 & {\bf EC 05089-5933} & 0.363\\
 5387935253342128256 &     162.99 &     -44.23  &    80.3 &   & \\
 5457346803926723456 &     161.68 &      -27.86 &      68.8&  &   \\
 5992085728570051840 &     246.34 &      -44.24 &       87.8&  &   \\
 6216887306090464000 &     219.62 &      -31.46 &      65.8 &  UCAC4 293-078484&  $0.19668$ \\
 6286158119187026688 &     218.94  &     -16.64 &      63.6 & {\bf HE 1432-1625} &  0.350$^{b}$\\
 6300000214666201472 &     216.58  &     -14.09 &      91.6 &    & \\
 6576232101460909056 &     321.67  &     -44.56 &      94.7 &  {\bf HE 2123-4446}&0.855\\
 674214551557961984  &     118.77  &      22.00 &      93.4 &  {\bf U Gem}  & 0.177$^{b}$\\
 783921244796958208  &     168.94  &      42.97  &      98.8&  {\bf AR UMa} &  0.080$^{b}$\\
 881086019353249280  &     119.83  &      32.33  &      91.5&  GALEX J075919.4+321948   & 0.259\\
 947545965334561280  &     106.44  &      39.58  &     89.6& [SGR2010] J0705+3934  & \\
\hline
\end{tabular}
\begin{tablenotes}
\item{a}: Name of WD or CV listed in SIMBAD and/or VSX. Bold font indicates the X-ray sources. 
\item{b}: Period listed in VSX.  
\end{tablenotes}
\end{table*}

\end{appendix}

\end{document}